\documentclass[10pt]{iopart}

\usepackage{amsfonts}
\usepackage{dsfont}
\usepackage{enumerate}

\newcommand{\prob}{\mathbb{P}}
\newcommand{\Ex}{\mathbb{E}}
\newcommand{\Rl}{\mathbb{R}}
\newcommand{\inter}{\mbox{int}\,}
\newcommand{\cl}{\mbox{cl}\,}
\newcommand{\ri}{\mbox{ri}\,}

\newcommand{\dom}{\mbox{dom}\,}

\newtheorem{lemma}{Lemma}
\newtheorem{proposition}{Proposition}
\newtheorem{theorem}{Theorem}
\newtheorem{assumption}{Assumption}

\begin{document}

\title[Large Deviations in Renewal Models of Statistical Mechanics]
      {Large Deviations in Renewal Models of Statistical Mechanics}

      \author{Marco Zamparo}
\address{Dipartimento Scienza Applicata e Tecnologia, Politecnico di Torino, Corso Duca degli Abruzzi 24,
  10129 Torino, Italy}
\ead{marco.zamparo@polito.it}

\begin{abstract}
In Ref. \cite{MarcoProb} the author has recently established sharp
large deviation principles for cumulative rewards associated with a
discrete-time renewal model, supposing that each renewal involves a
broad-sense reward taking values in a separable Banach space. The
renewal model has been there identified with constrained and
non-constrained pinning models of polymers, which amount to Gibbs
changes of measure of a classical renewal process.  In this paper we
show that the constrained pinning model is the common mathematical
structure to the Poland-Scheraga model of DNA denaturation and to some
relevant one-dimensional lattice models of Statistical Mechanics, such
as the Fisher-Felderhof model of fluids, the
Wako-Sait\^o-Mu\~noz-Eaton model of protein folding, and the
Tokar-Dreyss\'e model of strained epitaxy. Then, in the framework of
the constrained pinning model, we develop an analytical
characterization of the large deviation principles for cumulative
rewards corresponding to multivariate deterministic rewards that are
uniquely determined by, and at most of the order of magnitude of, the
time elapsed between consecutive renewals. In particular, we outline
the explicit calculation of the rate functions and successively we
identify the conditions that prevent them from being analytic and that
underlie affine stretches in their graphs. Finally, we apply the
general theory to the number of renewals. From the point of view of
Equilibrium Statistical Physics and Statistical Mechanics, cumulative
rewards of the above type are the extensive observables that enter the
thermodynamic description of the system. The number of renewals, which
turns out to be the commonly adopted order parameter for the
Poland-Scheraga model and for also the renewal models of Statistical
Mechanics, is one of these observables.\\

\end{abstract}

\maketitle

\section{Introduction}

{\it Renewal models} describe events that are randomly renewed over
time. Extensive use of renewal models as classical stochastic
processes is made in different areas of applied mathematics, including
Queueing Theory \cite{AsmussenBook}, Insurance \cite{DicksonBook}, and
Finance \cite{RSSTBook} among others. With a different interpretation
of the time coordinate, these models also enter Equilibrium
Statistical Physics through the phenomena of polymer pinning and
melting of DNA. Indeed, the thermodynamics of a polymer that is pinned
by a substrate at certain monomers regarded as renewed events along
the polymer chain is studied by a renewal model called the {\it
  pinning model} \cite{Giac,Hollander_Polymer}.  Similarly, DNA
denaturation upon heating has been investigated by Poland and Scheraga
\cite{PolSch1,PolSch2} through a renewal model where renewed events
identify base pairs along the DNA sequence. Formally, the
Poland-Scheraga model is a {\it constrained pinning model} obtained by
the pinning model under the condition that one of the renewals occurs
at a predetermined position corresponding to the DNA size
\cite{Giac}. Although it is generally not recognized, the constrained
pinning model also is the mathematical essence of some significant
one-dimensional lattice models of Statistical Mechanics.  They are the
cluster model of fluids proposed by Fisher and Felderhof
\cite{FisFel1,FisFel2,FisFel3,FisFel4,Roepstorff}, the model of
protein folding introduced independently by Wako and Sait\^o first
\cite{WaSa1,WaSa2} and Mu\~noz and Eaton later
\cite{MuEa1,MuEa2,MuEa3}, and the model of strained epitaxy considered
by Tokar and Dreyss\'e \cite{TokDre1,TokDre2,TokDre3}.  These models
have attracted the interest of many researchers due to exact
solvability, often encouraging generalizations such as in the case of
the Wako-Sait\^o-Mu\~noz-Eaton model
\cite{BruPe,ZaPe,BruPeZa,ImPeZa,ItSa1,ZaTroMa,ItSa2,Lee}.

In the framework of discrete-time renewal models, identified with
constrained and non-constrained homogeneous pinning models, the author
\cite{MarcoProb} has recently established large deviation principles
for cumulative rewards, supposing that each renewal involves a
broad-sense reward taking values in a separable Banach space.  {\it
  Deterministic rewards} that are uniquely determined by, and at most
of the order of magnitude of, the time elapsed between consecutive
renewals constitute a special class of rewards for which the theory
can be further developed in an analytical direction.  This class of
rewards deserves attention from the point of view of Equilibrium
Statistical Physics and Statistical Mechanics, because the
corresponding cumulative rewards are the extensive observables that
enter the thermodynamic description of the system.  So far, analytical
characterizations of large deviation principles for macroscopic
observables have been provided only for few lattice models of
Statistical Mechanics, including the Curie-Weiss model
\cite{Ellisbook}, the Curie-Weiss-Potts model \cite{CoElTou}, the
mean-field Blume-Emery-Griffiths model \cite{ElOtTou}, and the Ising
model to some extent \cite{El,FoOr,Ol,Geo}.

The present paper reconsiders the constrained pinning model as defined
in Ref. \cite{MarcoProb} with a dual purpose. First of all, it aims to
propose a unified formulation of the Poland-Scheraga model, the
Fisher-Felderhof model, the Wako-Sait\^o-Mu\~noz-Eaton model, and the
Tokar-Dreyss\'e model as a constrained pinning model.  The latter
three models are customarily presented in terms of binary occupation
numbers that are here interpreted as indicators of hypothetical
renewals, thus constituting the so-called {\it regenerative
  phenomenon} associated by Kingman with a renewal process
\cite{KingmanBook}.  To the best of our knowledge, the mapping of
these models with renewal systems has never been shown before.
Second, the paper aims to characterize analytically, within the
constrained pinning model, the rate functions associated with large
deviation principles for cumulative rewards corresponding to
multivariate deterministic rewards, thus providing a portrayal for the
(possible joint) fluctuations of macroscopic observables. In doing
this, the conditions that prevent the rate functions from being
analytic and that underlie affine stretches in their graphs are
identified.  The connection between the singular behavior of rate
functions and critical phenomena has been gaining considerable
interest in the physics community as demonstrated by several recent
works
\cite{Phys1,Phys2,Phys3,Phys4,Phys5,Phys6,Phys7,Phys8,Phys9,Phys10,Phys11,Phys12},
two of which dealing with renewal processes
\cite{Phys10,Phys11}. Renewal models supply a perfect framework to
probe this connection as they are able to account for phase
transitions of any order \cite{Giac}, making at the same time explicit
results feasible in contrast to most models of Statistical Mechanics.

The paper is organized as follows. In Sect.\ \ref{sec:pinning} we
introduce the framework of pinning models together with deterministic
rewards. In this section we also report, specialized to deterministic
rewards, the large deviation principle obtained in
Ref.\ \cite{MarcoProb} for constrained pinning models.  In
Sect.\ \ref{sec:statmech} we explain the role of the constrained
pinning model in Statistical Mechanics, briefly reviewing the
Fisher-Felderhof model, the Wako-Sait\^o-Mu\~noz-Eaton model, and the
Tokar-Dreyss\'e model as well as the Poland-Scheraga model. The rate
functions corresponding to the constrained pinning model are studied
in Sect.\ \ref{sec:I}, where their explicit calculation is outlined
and their main analytical properties are classified. Here we also
single out a {\it critical constrained pinning model} where persistent
large fluctuations of extensive observables lead to subexponential
decays of probabilities that cannot be captured by a large deviation
principle. An example concerning the number of renewals is finally
proposed to show how the analytical theory developed in the section
works in practice.  The major mathematical proofs are reported in the
appendices in order to not interrupt the flow of the presentation.

\section{Pinning Models, Deterministic Rewards, and Large Deviations}
\label{sec:pinning}

In this section we review the framework of pinning models as defined
in Ref. \cite{MarcoProb}.  Then, we introduce a class of deterministic
rewards in the Euclidean $d$-space $\Rl^d$ and, focusing on
constrained pinning models, we specialize to such class the large
deviation principle established in Ref. \cite{MarcoProb} for
cumulative rewards associated with general rewards in separable Banach
spaces.

\subsection{Pinning Models}

The pinning model considered in Ref. \cite{MarcoProb} calls for a
probability space $(\Omega,\mathcal{F},\prob)$ and random variables
$S_1,S_2,\ldots$ on it that take values in
$\{1,2,\ldots\}\cup\{\infty\}$ and form an independent and identically
distributed sequence. In the classical theory of renewal processes,
the variable $S_i$ is regarded as the {\it waiting time} for the $i$th
occurrence at the {\it renewal time} $T_i:=S_1+\cdots+S_i$ of some
event that is continuously renewed over time. Instead, here we imagine
that a polymer consisting of $t\ge 1$ monomers is pinned by a
substrate at the monomers $T_1,T_2,\ldots$ in such a way that the
monomer $T_i$ contributes an energy $-v(S_i)$ provided that $T_i\le
t$. The real function $v$ is called the {\it potential}. The state of
the polymer is described by the law $\prob_t$ defined on the
measurable space $(\Omega,\mathcal{F})$ by the Gibbs change of measure
\begin{eqnarray}
\nonumber
\frac{d\prob_t}{d\prob}:=\frac{e^{H_t}}{Z_t},
\end{eqnarray}
where $H_t:=\sum_{i\ge 1}v(S_i)\mathds{1}_{\{T_i\le t\}}$ is the {\it
  Hamiltonian} and the normalization constant $Z_t:=\Ex[e^{H_t}]$ is
the {\it partition function}. The model $(\Omega,\mathcal{F},\prob_t)$
precisely is the {\it pinning model}, that we supply with the
hypotheses of aperiodicity and extensivity.  The {\it waiting time
  distribution} $p:=\prob[S_1=\cdot\,]$ is said to be {\it aperiodic}
if its support $\mathcal{S}:=\{s\ge 1:p(s)>0\}$ is nonempty and there
does not exist an integer $\tau>1$ with the property that
$\mathcal{S}$ includes only some multiples of $\tau$. We observe that
$p$ can be made aperiodic by simply changing the time unit whenever
$\prob[S_1<\infty]>0$.
\begin{assumption}
\label{ass1}
The waiting time distribution $p$ is aperiodic.
\end{assumption}
The potential $v$ is said to be {\it extensive} if there exists a real
number $z_o$ such that $e^{v(s)}p(s)\le e^{z_os}$ for all
$s$. Extensivity is necessary to make the thermodynamic limit of the
pinning model meaningful since
$Z_t\ge\Ex[e^{H_t}\mathds{1}_{\{S_1=t\}}]=e^{v(t)}p(t)$.
\begin{assumption}
\label{ass2}
The potential $v$ is extensive.
\end{assumption}

This paper focuses on the {\it constrained pinning model} where the
last monomer is always pinned by the substrate. The constrained
pinning model as introduced in Ref. \cite{MarcoProb} corresponds to
the law $\prob_t^c$ defined on the measurable space
$(\Omega,\mathcal{F})$ through the change of measure
\begin{eqnarray}
\nonumber
  \frac{d\prob_t^c}{d\prob}:=\frac{U_t e^{H_t}}{Z_t^c},
\end{eqnarray}
$U_t:=\sum_{i\ge 1}\mathds{1}_{\{T_i=t\}}$ being the renewal indicator
that takes value 1 if $t$ is a renewal and value 0 otherwise, and
$Z_t^c:=\Ex[U_te^{H_t}]$ being the partition function. Aperiodicity of
the waiting time distribution gives $Z_t^c>0$ for all sufficiently
large $t$ \cite{MarcoProb}, thus ensuring that the constrained pinning
model is well-defined at least for~such~$t$.

\subsection{Deterministic Rewards and Large Deviation Principles}

The {\it cumulative reward} by the integer time $t$ is
$W_t:=\sum_{i\ge 1}X_i\mathds{1}_{\{T_i\le t\}}$, supposing that the
$i$th renewal involves a {\it reward} $X_i$ valued in a vector space
and possibly dependent on $S_i$. Notice that $W_t$ reduces to the
number $N_t:=\sum_{\tau=1}^tU_\tau$ of renewals by $t$ when $X_i=1$
for all $i$.  The large deviation theory developed in
Ref. \cite{MarcoProb} describes the fluctuations of $W_t$ within
constrained and non-constrained pinning models for rewards that are
generic random variables valued in a real separable Banach space. In
this paper we deepen the study for the special case of {\it
  deterministic rewards} of the form $X_i:=f(S_i)$ for each $i$, where
$f$ is a function on $\{1,2,\ldots\}\cup\{\infty\}$ that takes values
in the Euclidean $d$-space $\Rl^d$ and satisfies the following
assumption.
\begin{assumption}
  \label{ass3}
If the support $\mathcal{S}$ of the waiting time distribution is
infinite, then $f(s)/s$ has a limit $r\in\Rl^d$ when $s$ goes to
infinity through $\mathcal{S}$.
\end{assumption}
Under this assumption, there exists a positive constant $M<\infty$
such that $\|f(s)\|\le Ms$ for every $s\in\mathcal{S}$, meaning that
$f$ is at most of the order of magnitude of the waiting time.  From
now on, $u\cdot v$ denotes the usual dot product between $u$ and $v$
in $\Rl^d$ and $\|u\|:=\sqrt{u\cdot u}$ is the Euclidean norm of $u$.

The large deviation principle for cumulative rewards in constrained
pinning models stated by theorems 1 of Ref. \cite{MarcoProb} can be
specialized to deterministic rewards as follows. With reference to the
formalism of \cite{MarcoProb}, it is convenient here to identify a
linear functional $\varphi$ on $\Rl^d$ with that unique $k\in\Rl^d$
such that $\varphi(w)=k\cdot w$ for all $w$. Let $z$ be the function
that maps each point $k\in\Rl^d$ in the extended real number $z(k)$
defined by
\begin{equation}
z(k):=\inf\bigg\{\zeta\in\Rl\,:\,\sum_{s\ge 1}e^{k\,\cdot f(s)+v(s)-\zeta s}\,p(s)\le 1\bigg\},
\label{defzetafun}
\end{equation}
where the infimum over the empty set is customarily interpreted as
$\infty$. Denote by $I$ the Fenchel-Legendre transform of $z-z(0)$,
which associates every vector $w\in\Rl^d$ with the extended real
number $I(w)$ defined by
\begin{equation}
I(w):=\sup_{k\in\Rl^d}\Big\{w\cdot k-z(k)+z(0)\Big\}.
\label{defIorate}
\end{equation}
We point out that the function $z$ is finite everywhere under
assumptions \ref{ass1}, \ref{ass2}, and \ref{ass3}. Indeed, given any
$k\in\Rl^d$, assumption \ref{ass1} entailing $\sum_{s\ge 1}p(s)>0$
yields $\sum_{s\ge 1}e^{k\,\cdot f(s)+v(s)-\zeta s}\,p(s)>1$ for all
sufficiently negative $\zeta$, so that $z(k)>-\infty$. At the same
time, the bounds $e^{v(s)}p(s)\le e^{z_os}$ for each $s$ with some
real number $z_o$ by assumption \ref{ass2} and $\|f(s)\|\le M s$ for
every $s\in\mathcal{S}$ with some constant $M<\infty$ by assumption
\ref{ass3} give $\sum_{s\ge 1}e^{k\cdot f(s)+v(s)-\zeta s}\,p(s)\le 1$
for all $\zeta\ge z_o+M\|k\|+\ln2 $, thus implying $z(k)\le
z_o+M\|k\|+\ln2<\infty$. The finiteness of $z$ allows us to obtain the
following strong version of theorem 1 of Ref. \cite{MarcoProb}, which
extends the Cram\'er's theorem to the cumulative reward $W_t$ within
the constrained pinning model $(\Omega,\mathcal{F},\prob_t^c)$.
\begin{theorem}
\label{mainth1}
  The following conclusions hold under assumptions \ref{ass1},
  \ref{ass2}, and \ref{ass3}:
  \begin{enumerate}[{\upshape(a)}]
  \setlength\itemsep{0.3em}
  \item the function $z$ is finite everywhere and convex. The function
    $I$ is lower semicontinuous and proper convex;
  \item if $G\subseteq\Rl^d$ is an open set, then
    \begin{eqnarray}
    \nonumber
      \liminf_{t\uparrow\infty}\frac{1}{t}\ln\prob_t^c\bigg[\frac{W_t}{t}\in G\bigg]\ge -\inf_{w\in G}\{I(w)\};
    \end{eqnarray}
    \setlength\itemsep{0em}
\item if $F\subseteq\Rl^d$ is either a closed set or a Borel convex set, then
  \begin{eqnarray}
  \nonumber
      \limsup_{t\uparrow\infty}\frac{1}{t}\ln\prob_t^c\bigg[\frac{W_t}{t}\in F\bigg]\le-\inf_{w\in F}\{I(w)\}.
\end{eqnarray}
\end{enumerate}
\end{theorem}

The lower bound in part (b) and the upper bound in part (c) are
called, respectively, {\it large deviation lower bound} and {\it large
  deviation upper bound} \cite{DemboBook,Hollander}. When a lower
semicontinuous function $I$ exists so that the large deviation lower
bound holds for each open set $G$ and the large deviation upper bound
holds for each closed set $F$, then $W_t$ is said to satisfy a {\it
  large deviation principle} with {\it rate function} $I$
\cite{DemboBook,Hollander}. Theorem \ref{mainth1} states that the
cumulative reward $W_t$ satisfies a large deviation principle with
rate function $I$ given by (\ref{defIorate}) within the constrained
pinning model.  We observe that the rate function $I$ has compact
level sets, thus resulting in a {\it good rate function}
\cite{DemboBook,Hollander}. Indeed, the level set $\{w\in\Rl^d:I(w)\le
a\}$ for a given positive real number $a$ is closed by the lower
semicontinuity of $I$ and bounded as if $w\ne 0$ belongs to this set,
then the bounds $w\cdot k-z(k)+z(0)\le I(w)\le a$ and $z(k)\le
z_o+M\|k\|+\ln2$ together imply $\|w\|\le z_o+M+\ln2-z(0)$ if the
choice $k:=w/\|w\|$ is made.

\section{Uses of the Constrained Pinning Model in Statistical Mechanics}
\label{sec:statmech}

This section resolves around the binary process $\{U_t\}_{t\ge 0}$ of
renewal indicators, with $U_0:=1$. We recall that $U_t:=1$ if $t$ is a
renewal and $U_t:=0$ otherwise for each $t\ge 1$. From a mathematical
point of view, the finite-dimensional marginals of the process
$\{U_t\}_{t\ge 0}$ with respect to the constrained pinning model
coincide with the finite-volume Gibbs state associated with the
Fisher-Felderhof model of fluids, the Wako-Sait\^o-Mu\~noz-Eaton model
of protein folding, and the Tokar-Dreyss\'e model of strained
epitaxy. Here we determine the finite-dimensional marginals of
$\{U_t\}_{t\ge 0}$ with respect to constrained and non-constrained
pinning models. Then, we briefly review the above models and the
Poland-Scheraga model, sketching the mapping with the constrained
pinning model.

\subsection{Kingman's Regenerative Phenomena and Pinning Models}

The binary process $\{U_t\}_{t\ge 0}$ is a {\it discrete-time
  regenerative phenomenon} according to Kingman \cite{KingmanBook},
because it satisfies the following property. This property is proved
in \ref{proof_Kingman} and comes from the fact that a renewal process
forgets the past and starts over at every renewal.
\begin{proposition}
\label{King}
For any $m\ge 1$ and instants $0=:\tau_0<\tau_1<\cdots<\tau_m$
\begin{eqnarray}
\nonumber
\prob\big[U_{\tau_1}=\cdots=U_{\tau_m}=1\big]=\prod_{l=1}^m\prob\big[U_{\tau_l-\tau_{l-1}}=1\big].
\end{eqnarray}
\end{proposition}

The finite-dimensional marginals of the process $\{U_t\}_{t\ge 0}$
with respect to constrained and non-constrained pinning models can be
determined through the following argument. Fix a time $t\ge 1$ and
binary numbers $u_1,\ldots,u_t$ that are supposed to contain
$n:=\sum_{\tau=1}^tu_\tau\ge 1$ ones in certain positions, the $i$th
of which being written as $s_1+\cdots+s_i$. The distance $s\le t$
between consecutive ones is attained a number of times equal to
$\sum_{i=1}^n\mathds{1}_{\{s_i=s\}}$, which can be explicitly
expressed in terms of $u_1,\ldots,u_t$ and $u_0=1$ as
\begin{equation}
\#_{s|t}(u_0,\ldots,u_t):=\sum_{\tau=1}^{t-s+1}u_{\tau-1}\Bigg[\prod_{k=\tau}^{\tau+s-2}(1-u_k)\Bigg]u_{\tau+s-1},
\label{numbers}
\end{equation}
where the intermediate factor is not present when $s=1$. The distance
between the position of the last one and $t$ is
$t-s_1-\cdots-s_n=\sum_{\tau=1}^t\prod_{k=\tau}^t(1-u_k)$. The
condition $U_\tau=u_\tau$ for each $\tau\le t$ is tantamount to the
condition $S_i=s_i$ for each $i\le n$ and $S_{n+1}>t-t_n$ provided
that $u_0=1$ since $U_0:=1$.  It follows that
\begin{eqnarray}
\nonumber
\prob_t\big[U_0=u_0,\ldots,U_t=u_t\big]&=&\int_\Omega d\prob_t\prod_{\tau=0}^t\mathds{1}_{\{U_\tau=u_\tau\}} \\
  \nonumber
  &=&\frac{1}{Z_t}\int_\Omega d\prob\prod_{\tau=0}^t\mathds{1}_{\{U_\tau=u_\tau\}} \,e^{H_t} \\
  \nonumber
  &=&\frac{u_0}{Z_t}\int_\Omega d\prob\prod_{i=1}^n\mathds{1}_{\{S_i=s_i\}} \,\mathds{1}_{\{S_{n+1}>t-s_1-\cdots-s_n\}} \, e^{\sum_{i=1}^nv(S_i)} \\
  \nonumber
  &=&\frac{u_0}{Z_t}\prod_{i=1}^ne^{v(s_i)}\,\prob\big[S_1=s_i\big] \, \prob\big[S_1>t-s_1-\cdots-s_n\big]\\
\nonumber
&=&\frac{u_0}{Z_t}\prod_{s=1}^t\big[e^{v(s)}p(s)\big]^{\sum_{i=1}^n\mathds{1}_{\{s_i=s\}}}\,\prob\big[S_1>t-s_1-\cdots-s_n\big]\\
\nonumber
&=&\frac{u_0}{Z_t}\prod_{s=1}^t\big[e^{v(s)}p(s)\big]^{\#_{s|t}(u_0,\ldots,u_t)}\,\prob\Bigg[S_1>\sum_{\tau=1}^t\prod_{k=\tau}^t(1-u_k)\Bigg].
\end{eqnarray}
This formula also holds for $n=0$, which corresponds to the case
$u_1=\cdots=u_t=0$ that gives $\#_{s|t}(u_0,\ldots,u_t)=0$ for all
$s$, since the probability that $U_1=\cdots=U_t=0$ is $\prob[S_1>t]$.
This way, we find that the finite-dimensional marginals of the process
$\{U_t\}_{t\ge 0}$ with respect to the pinning model are expressed for
every integer $t\ge 1$ and binary numbers $u_0,\ldots,u_t$ by
\begin{eqnarray}
\nonumber
  \prob_t\big[U_0=u_0,\ldots,U_t=u_t\big]=\frac{u_0}{Z_t}\prod_{s=1}^t\big[e^{v(s)}p(s)\big]^{\#_{s|t}(u_0,\ldots,u_t)}\,
  \prob\Bigg[S_1>\sum_{\tau=1}^t\prod_{k=\tau}^t(1-u_k)\Bigg].
\end{eqnarray}
As far as the constrained pinning model is concerned, adding the
condition $U_t=1$ in this expression we get
\begin{equation}
\prob_t^c\big[U_0=u_0,\ldots,U_t=u_t\big]=\frac{u_0u_t}{Z_t^c}\prod_{s=1}^t \big[e^{v(s)}p(s)\big]^{\#_{s|t}(u_0,\ldots,u_t)}.
\label{distr_constr_X}
\end{equation}
The corresponding law for waiting times is obtained by noticing that
$n$ renewals occur by the time $t\ge 1$, namely $N_t=n$, and a renewal
exactly occurs at the time $t$ if and only if
$T_n=\sum_{i=1}^nS_i=t$. This argument gives for every positive
integers $n$ and $s_1,\ldots,s_n$ the formula
\begin{equation}
\prob_t^c\big[S_1=s_1,\ldots,S_n=s_n,N_t=n\big]=\frac{\mathds{1}_{\{\sum_{i=1}^ns_i=t\}}}{Z_t^c}\prod_{i=1}^ne^{v(s_i)}p(s_i).
\label{distr_constr_S}
\end{equation}

The probability distribution (\ref{distr_constr_X}) is exactly the
finite-volume Gibbs state associated with the Fisher-Felderhof model,
the Wako-Sait\^o-Mu\~noz-Eaton model, and the Tokar-Dreyss\'e model,
whereas the probability distribution (\ref{distr_constr_S}) is the
Poland-Scheraga model. The following four paragraphs illustrate such
connections.

\subsection{The Model by Poland and Scheraga for Melting of DNA}

Most DNA molecules consist of two strands made up of nucleotide
monomers.  Monomers on one strand are bound to a specific matching
monomer on the other strand, constituting the so-called Watson-Crick
pairs that held together the two strands in a double helix.  Thermal
denaturation of DNA is the process by which the two strands unravel
upon heating, melting into bubbles where they are apart. The formation
of a bubble results in an entropic gain $\sigma_l$, which is observed
experimentally to depend on the length $l$ of the denatured fragment
as $\sigma_l\sim al+b-c\ln l$ with positive coefficients $a$, $b$, and
$c$ \cite{PolSch2}.  The logarithmic dependence can be explained as a
consequence of loop closure since the bubble can be regarded as a loop
of length $2l$ \cite{PolSch2}.  The Poland-Scheraga model is a
simplified model that aims to describe thermal denaturation of DNA as
a phase transition \cite{PolSch1,PolSch2}. The model considers a
partially melted DNA molecule as being composed of an alternating
sequence of bound segments and denaturated segments that do not
interact with one another. A bound segment of length $l\ge 1$ is
favored by the energetic gain $\epsilon l$, the binding energy
$\epsilon<0$ being taken to be the same for all matching monomers,
whereas a denaturated segment of length $l\ge 1$ is favored by a
certain entropic gain $\sigma_l>0$.  Here we suppose that $\epsilon$
is measured in unit of $k_BT$, where $k_B$ is the Boltzmann constant
and $T$ is the absolute temperature. The mathematical construction of
the Poland-Scheraga model with $t$ monomers per strand assumes that
there is a variable number $n$ of consecutive stretches of positive
lengths $s_1,\ldots,s_n$ that span a chain of $t$ monomers: $1\le n\le
t$ and $\sum_{i=1}^ns_i=t$. The $i$th stretch is imagined to consist
of just one bound monomer if $s_i=1$ and one denaturated segment of
length $s_i-1$ followed by one bound monomer if $s_i>1$. Within this
scheme, monomers $s_1,s_1+s_2,\ldots,s_1+\cdots +s_n$ are bound and
there is actually a bound segment of length $l\ge 2$ starting at
position $s_1+\cdots+s_i$ if $s_i>1$ when $i>1$,
$s_{i+1}=\cdots=s_{i+l-1}=1$, and $s_{i+l}>1$ when $i+l\le n$.  We
notice that the last monomer is always bound under this construction.
The probability that the Poland-Scheraga model assigns to a
configuration with $n$ stretches of lengths $s_1,\ldots,s_n$ reads
\begin{equation}
{\rm PS}_t(n;s_1,\ldots,s_n):=\frac{1}{\Xi_t}\prod_{i=1}^n\exp\big(-\epsilon+\sigma_{s_i-1}\big),
  \label{Gibbs_PL}
\end{equation}
where $\sigma_0:=0$ and $\Xi_t$ is the partition function:
\begin{eqnarray}
\nonumber
\Xi_t:=\sum_{n=1}^t\sum_{s_1\ge 1}\cdots\sum_{s_n\ge 1}\mathds{1}_{\{\sum_{i=1}^ns_i=t\}}\prod_{i=1}^n\exp\big(-\epsilon+\sigma_{s_i-1}\big).
\end{eqnarray}

A similarity between the probability distributions (\ref{Gibbs_PL})
and (\ref{distr_constr_S}) is evident and can be made tight as
follows.  Focusing on the configuration with only one large denatured
segment we get $\ln \Xi_t\ge -\epsilon+\sigma_{t-1}$, so that a
necessary condition for the free energy $\ln \Xi_t$ to be extensive in
$t$ is $\eta_o:=\limsup_{l\uparrow\infty}\sigma_l/l<\infty$. Under
this condition, a real number $\eta\ge\eta_o$ can be found in such a
way that $\sum_{s\ge 1}p(s)\le 1$ with $p(s):=e^{\sigma_{s-1}-\eta s}$
for each $s$ and a constrained pinning model with waiting time
distribution $p$ and potential $v(s):=-\epsilon$ for every $s$ can be
devised. The distribution $p$ is clearly aperiodic because $p(s)>0$
for all $s$. Such constrained pinning model gives rise to the
identities ${\rm
  PS}_t(n;s_1,\ldots,s_n)=\prob_t^c[S_1=s_1,\ldots,S_n=s_n,N_t=n]$
whenever $\sum_{i=1}^ns_i=t$ and $\ln\Xi_t=\ln Z_t^c+\eta t$.  Thus,
we get an interpretation of the Poland-Scheraga model as a constrained
pinning model where renewal times mark bound monomers. It is worth
noting here that if $\sum_{s\ge 1}e^{\sigma_{s-1}-\eta_o s}\ge 1$,
then $\eta$ can be chosen in such a way that $\sum_{s\ge 1}p(s)=1$,
giving $\prob[S_1=\infty]=0$. If on the contrary $\sum_{s\ge
  1}e^{\sigma_{s-1}-\eta_o s}<1$, then any choice of $\eta$ entails
$\prob[S_1=\infty]>0$.

Extensive observables involved in the thermodynamic description of the
system are, for example, the number $N_t$ of bound monomers per strand
and the total loop entropy. They are the cumulative rewards $W_t$
corresponding to the deterministic rewards $X_i:=f(S_i)$ for each $i$
with, respectively, $f(s)=1$ and $f(s)=\sigma_{s-1}$ for any $s$.  The
joint fluctuations of the number of bound monomers and the total loop
entropy can be investigated by taking $f(s)=(1,\sigma_{s-1})$ for all
$s$. As an application of the analytical theory developed in
Sect.\ \ref{sec:I}, in Par.\ \ref{application} we shall review the
phase transition of the Poland-Scheraga model and investigate the
fluctuations of $N_t$.  In particular, we will see that there are
situations where the fraction of bound monomers changes from being
positive to being zero at a certain finite value of the binding energy
while increasing $\epsilon$.

\subsection{The Model by Fisher and Felderhof for Fluids}

In 1970 Fisher and Felderhof published a series of papers where they
introduced a many-body cluster interaction model of a one-dimensional
continuum classical fluid \cite{FisFel1,FisFel2,FisFel3,FisFel4}. In
the thermodynamic limit, the model was found to exhibit a phase
transition from a gas-like phase containing clusters of particles of
all sizes to a liquid-like phase consisting essentially of a single
macroscopic cluster \cite{FisFel1,FisFel2}. The discrete counterpart
was later considered by Roepstorff \cite{Roepstorff}, who formalized a
lattice version of the model where, in a nutshell, if some site is not
occupied by a particle, then the particles on the left do not interact
with those on the right. This means that particles interact only when
they fill a cluster of contiguous sites, contributing a certain energy
$E_l<0$ when the cluster has size $l\ge 1$. The model by Fisher and
Felderhof on a lattice is a lattice-gas model that can be introduced
as follows by taking advantage of (\ref{numbers}).  If particles are
arranged on $t$ lattice sites and if the binary variable $u_\tau$ is
associated with the site $\tau$ in such a way that $u_\tau=1$ denotes
a hole and $u_\tau=0$ denotes a particle, then
$\#_{s|t}(1,u_1,\ldots,u_t)$ defined by (\ref{numbers}) counts the
number of clusters with $s-1$ particles provided that $u_t=1$.  This
way, assuming that the last site is always a hole, namely $u_t=1$, the
probability that the Fisher-Felderhof model with a chemical potential
$\mu$ assigns to the configuration $u_1,\ldots,u_t$ can written as
\begin{equation}
{\rm FF}_t(u_1,\ldots,u_t):=\frac{u_t}{\Xi_t}\exp\Bigg[-\sum_{s=2}^t E_{s-1}\,\#_{s|t}(1,u_1,\ldots,u_t)+\mu\sum_{\tau=1}^t(1-u_\tau)\Bigg],
\label{Gibbs0_FF}
\end{equation}
where $\Xi_t$ is the partition function:
\begin{eqnarray}
\nonumber
  \Xi_t:=\sum_{u_1=0}^1\cdots\sum_{u_t=0}^1
  u_t\exp\Bigg[-\sum_{s=2}^t E_{s-1}\,\#_{s|t}(1,u_1,\ldots,u_t)+\mu\sum_{\tau=1}^t(1-u_\tau)\Bigg].
\end{eqnarray}
Here parameters are supposed to be expressed in unit of $k_BT$.

From the mathematical point of view, the probability distribution
(\ref{Gibbs0_FF}) is nothing but (\ref{distr_constr_X}). This fact is
understood by observing at first that the identities
$\sum_{s=1}^t\#_{s|t}(1,u_1,\ldots,u_t)=\sum_{\tau=1}^tu_\tau$ and
$\sum_{s=1}^ts\,\#_{s|t}(1,u_1,\ldots,u_t)=t$, which are valid when
$u_t=1$, allow us to recast (\ref{Gibbs0_FF}) in the form
\begin{eqnarray}
\nonumber
{\rm FF}_t(u_1,\ldots,u_t):=\frac{u_te^{(\mu-\eta)t}}{\Xi_t}\exp\Bigg[\sum_{s=1}^t\big(-\mu+\eta s-E_{s-1}\big)\#_{s|t}(1,u_1,\ldots,u_t)\Bigg],
\end{eqnarray}
where the convention $E_0:=0$ has been made and an arbitrary number
$\eta$ has been introduced. Second, let us notice that a necessary
condition for the free energy $\ln \Xi_t$ to be extensive in $t$ is
$\eta_0:=\liminf_{l\uparrow\infty}E_l/l>-\infty$. Under this
condition, a real number $\eta\le\eta_o$ exists with the property that
$\sum_{s\ge 1}p(s)\le 1$ with $p(s):=e^{\eta s-E_{s-1}}$ for each
$s\ge 1$. Then, the constrained pinning model with aperiodic waiting
time distribution $p$ and potential $v(s):=-\mu$ for every $s$
satisfies ${\rm
  FF}_t(u_1,\ldots,u_t)=\prob_t^c[U_0=u_0,\ldots,U_t=u_t]$ whenever
$u_0=1$ and $\ln\Xi_t=\ln Z_t^c+(\mu-\eta)t$.  This way, the
Fisher-Felderhof model can be interpreted as a constrained pinning
model where renewal times mark holes. As for the Poland-Scheraga
model, $\eta$ can be chosen so that $\prob[S_1=\infty]=0$ only when
$\sum_{s\ge 1}e^{\eta_os-E_{s-1}}\ge 1$.  The number $N_t$ of holes
and the total energy are extensive observables entering the
thermodynamic description of the system, the latter being the
cumulative reward $W_t$ associated with the deterministic rewards
$X_i:=E_{S_i-1}$ for every $i$. Similarly to the Poland-Scheraga
model, the Fisher-Felderhof model can have a gas-liquid phase
transition. We shall review this phase transition in
Par.\ \ref{application}, showing that there are situations where the
fraction of holes changes from being positive to being zero at a
certain finite value of the chemical potential while increasing $\mu$.

\subsection{The Model by Wako, Sait\^o, Mu\~noz, and Eaton for Protein Folding}

Most proteins consist of a long chain of amino acid monomers held
together by peptide bonds.  At physiological temperatures, peptide
bonds are planar, rigid, covalent bonds that allow adjacent monomers
to perform two rotations. Native values of the dihedral angles
associated with these two rotations identify the functional
three-dimensional structure of the protein.  Protein folding is the
cooperative process by which a polypeptide chain folds into its native
shape from random coil.  The model by Wako and Sait\^o
\cite{WaSa1,WaSa2} and Mu\~noz and Eaton \cite{MuEa1,MuEa2,MuEa3} is a
simplified discrete model that aims to describe the process of protein
folding as a first-order-like phase transition. This model considers a
protein made up of $t+1$ monomers as a sequence of $t$ peptide
bonds. A configuration of the protein is identified by associating the
$i$th peptide bond with a binary variable $u_i$ taking value 0 if the
dihedral angles are native and value 1 otherwise. Bonds $i$ and $j>i$
are supposed to interact only if all intervening bonds along the chain
are native, namely only if $u_i=\cdots=u_j=0$. For homogeneous systems
like homopolymers \cite{WaSa1}, their interaction contributes the
energy $\epsilon_{j-i}\le 0$, which we express here in unit of
$k_BT$. In order to incorporate the principle of minimal frustration
of proteins, the model imposes the condition $\epsilon_{j-i}=0$ unless
bonds $i$ and $j$ are known to be in spatial proximity in the native
three-dimensional structure of the protein.  The model also takes into
account the entropic loss $\sigma>0$ of fixing one peptide unit in the
native conformation \cite{WaSa1,MuEa3}. Assuming for our convenience
that the last bond is always disordered, so that $u_t=1$, the
probability that the Wako-Sait\^o-Mu\~noz-Eaton model assigns to the
configuration $u_1,\ldots,u_t$ reads
\begin{equation} 
  {\rm WSME}_t(u_1,\ldots,u_t):=\frac{u_t}{\Xi_t}
  \exp\Bigg[-\sum_{i=1}^{t-1}\sum_{j=i+1}^t \epsilon_{j-i}\prod_{k=i}^j (1-u_k)+\sigma\sum_{i=1}^tu_i\Bigg],
\label{Gibbs0_WSME}
\end{equation}
where $\Xi_t$ is the partition function:
\begin{eqnarray}
\nonumber
  \Xi_t:=\sum_{u_1=0}^1\cdots\sum_{u_t=0}^1
  u_t\exp\Bigg[-\sum_{i=1}^{t-1}\sum_{j=i+1}^t \epsilon_{j-i}\prod_{k=i}^j (1-u_k)+\sigma\sum_{i=1}^tu_i\Bigg].
\end{eqnarray}

The probability distribution (\ref{Gibbs0_WSME}) can be recast in the
form (\ref{distr_constr_X}) by making the number
$\#_{s|t}(1,u_1,\ldots,u_t)$ defined by (\ref{numbers}) appear. To
this aim, set $E_0:=0$ and $E_l:=\sum_{s=1}^l(l-s)\,\epsilon_s$ for
each $l\ge 1$. We notice that $E_l$ is the energetic contribution of a
stretch of $l$ consecutive native bonds.  Then, the three simple
identities $\sum_{s=1}^t
E_{s-1}\,\#_{s|t}(1,u_1,\ldots,u_t)=\sum_{i=1}^{t-1}\sum_{j=i+1}^t\epsilon_{j-i}\prod_{k=i}^j(1-u_k)$,
$\sum_{s=1}^t\#_{s|t}(1,u_1,\ldots,u_t)=\sum_{i=1}^tu_i$, and
$\sum_{s=1}^ts\,\#_{s|t}(1,u_1,\ldots,u_t)=t$, which hold when
$u_t=1$, allow us to introduce an arbitrary number $\eta$ and rewrite
(\ref{Gibbs0_WSME}) as
\begin{eqnarray}
\nonumber
{\rm WSME}_t(u_1,\ldots,u_t)=\frac{u_te^{-\eta t}}{\Xi_t}\exp\Bigg[\sum_{s=1}^t\big(\sigma+\eta s-E_{s-1}\big)\#_{s|t}(1,u_1,\ldots,u_t)\Bigg].
\end{eqnarray}
We exploit $\eta$ to define a waiting time distribution. A necessary
condition for the free energy $\ln \Xi_t$ to be extensive in $t$ is
$\eta_0:=\liminf_{l\uparrow\infty}E_l/l>-\infty$.  Under this
condition, we can repeat the arguments made above for the
Fisher-Felderhof model to conclude that $\eta\le\eta_o$ exists so that
the aperiodic waiting time distribution $p$ and the potential $v$
defined by $p(s):=e^{\eta s-E_{s-1}}$ and $v(s):=\sigma$ for any $s$
originate a constrained pinning model whose marginal distribution
fulfills ${\rm
  WSME}_t(u_1,\ldots,u_t)=\prob_t^c[U_0=u_0,\ldots,U_t=u_t]$ if
$u_0=1$ and $\ln\Xi_t=\ln Z_t^c-\eta t$.  Thus, the
Wako-Sait\^o-Mu\~noz-Eaton model results in a constrained pinning
model where renewal times mark peptide bonds that do not take their
native conformation. The number of these bonds is the extensive
observable $N_t$. Another extensive observable is the total energy,
which is the cumulative reward $W_t$ associated with the deterministic
rewards $X_i:=E_{S_i-1}$ for every $i$. In Par.\ \ref{application} we
shall see that, while decreasing $\sigma$ or the energetic
contributions $E_l$, the model can exhibit a phase transition from a
denaturated phase to the native state where the fraction of native
bonds is 1.

\subsection{The Model by Tokar and Dreyss\'e for Strained Epitaxy}

Epitaxy is the growth process of a crystal film on a crystalline
substrate used in nanotechnology and in semiconductor fabrication.  In
most cases where the film material is different from the substrate
material, the strain of the crystal film to accommodate the lattice
geometry of the substrate leads to the self-assembly of coherent
nanostructures. The model by Tokar and Dreyss\'e
\cite{TokDre1,TokDre2,TokDre3} for strained epitaxy is a simplified
lattice-gas model that aims to describe the size distribution of these
atomic structures assuming that atoms interact effectively only when
they belong to the same cluster.  In the one-dimensional case,
clusters of $l\ge 1$ contiguous atoms contribute the energy $E_l<0$
\cite{TokDre1,TokDre2,TokDre3}.  When atoms are arranged on $t$
lattice sites and the configuration of the system is described by
binary variables taking value 1 for holes and value 0 for atoms, then
the Tokar-Dreyss\'e model endowed with a chemical potential $\mu$
formally is the Fisher-Felderhof model.  This way, the Tokar-Dreyss\'e
model can be identified with a constrained pinning model associated
with an aperiodic waiting time distribution under the condition
$\liminf_{l\uparrow\infty}E_l/l>-\infty$.

\section{Rate Functions Within the Constrained Pinning Model}
\label{sec:I}

This section addresses the study of the rate function $I$ defined by
(\ref{defIorate}) under assumptions \ref{ass1}, \ref{ass2}, and
\ref{ass3}, which will be made from now on.  The computation of $I$ by
means of methods from convex analysis is discussed first. Second, we
classify the main analytical properties of $I$, identifying the
conditions that prevent $I$ from being analytic and connecting the
presence of affine stretches in its graph with the existence of some
point where the function $z$ given by (\ref{defzetafun}) is not
differentiable. Third, we show that the large deviation principle
stated by theorem \ref{mainth1} loses its effectiveness to describe
the way the probabilities of the cumulative reward decay when the
origin is a point where $z$ is not differentiable. This fact leads us
to define the {\it critical constrained pinning model} where the decay
of probabilities is subexponential.  Finally, we exemplify the
analytical theory developed in this section by describing the
fluctuations of the number $N_t:=\sum_{\tau=1}^tU_\tau$ of renewals by
$t$ in the Poland-Scheraga model and in the other renewal models of
Statistical Mechanics.  Within these models, the extensive observable
$N_t$ is commonly regarded as a natural order parameter in order to
identify a phase transition.

Hereafter, we denote by $\inter A$ and $\cl A$ the interior and the
closure, respectively, of a set $A$ in $\Rl^d$.  The interior which
results when $A$ is regarded as a subset of its affine hull is the
relative interior, denoted by $\ri A$. Clearly, $\ri A=\inter A$ if
$A$ is of full dimension, i.e. has the whole space $\Rl^d$ as its
affine hull. A function $\varphi$ defined on an open set
$A\subseteq\Rl^d$ is analytic on $A$ if it can be represented by a
convergent power series in some neighborhood of any point $x\in A$. A
vector field $\nu$ on $A$ is analytic on $A$ if each of its components
is analytic on $A$.  Given a convex function $\varphi$ on $\Rl^d$, we
denote by $\dom \varphi:=\{x\in\Rl^d:\varphi(x)<\infty\}$ its
effective domain and by $\partial \varphi(x):=\{g\in\Rl^d:g\mbox{ is a
  subgradient of $\varphi$ at $x$}\}$ its subdifferential at $x$.  The
main results from convex analysis that will be used in the sequel are
recalled in \ref{convex}.

\subsection{Computing the Rate Function $I$}

In principle, the computation of the rate function $I$ is feasible
once the effective domain of $I$ and the subdifferentials of $z$ are
known. This can be deduced from the following proposition, which
collects together standard results from convex analysis (see
\ref{convex}, propositions \ref{riC}, \ref{riC_phi}, \ref{sub_FL}, and
\ref{domain_FL}). We point out that such results apply because $z$ and
$I$ are proper convex function by theorem \ref{mainth1}, $z$ being
continuous and $I$ being lower-semicontinuous. Continuity of $z$ is a
consequence of the fact that $z$ is a convex function that is finite
on the whole space $\Rl^d$.
\begin{proposition}
\label{compute_Io}
The following conclusions hold:
\begin{enumerate}[{\upshape(a)}]
  \setlength\itemsep{0.3em}
\item $I(w)=w\cdot k-z(k)+z(0)$ for every points $w$ and $k$ in
  $\Rl^d$ such that $w\in\partial z(k)$;
\item $w\in\partial z(k)$ if and only if $k\in\partial I(w)$;
\item for any {\upshape $w\in\ri(\dom I)$} there exists $k\in\Rl^d$
  with the property that $w\in\partial z(k)$;
\item for each {\upshape $w\in\cl(\dom I)$}, {\upshape $u\in\ri(\dom
  I)$}, and $\lambda\in[0,1)$ the vector $\lambda w+(1-\lambda)u$
  belongs to {\upshape $\ri(\dom I)$} and $I(w)=\lim_{\lambda\uparrow
    1}I(\lambda w+(1-\lambda)u)$.
\end{enumerate}
\end{proposition}
 
We are thus led to investigate the effective domain of $I$ and the
subdifferentials of $z$.  Let $\mathcal{S}:=\{s\ge 1:p(s)>0\}$ be the
support of the waiting time distribution $p$ as defined in
Sect. \ref{sec:pinning} and recall that $f$ denotes the function that
identifies deterministic rewards. The set $\mathcal{C}$ of all convex
combinations of the elements from $\{f(s)/s\}_{s\in\mathcal{S}}$ is
the smallest convex set that contains $\{f(s)/s\}_{s\in\mathcal{S}}$,
namely its convex hull.  The interest in the set $\mathcal{C}$ stems
from the fact that the effective domain of $I$ differs very little
from $\mathcal{C}$, as stated by the next proposition that is proved
in \ref{lemmadom}. In particular, the proposition entails $\cl(\dom
I)=\cl\mathcal{C}$ and $\ri(\dom I)=\ri\mathcal{C}$.
\begin{proposition}
\label{domIo}
Let $\mathcal{C}$ be the convex hull of
$\{f(s)/s\}_{s\in\mathcal{S}}$. Then, {\upshape
  $\mathcal{C}\subseteq\dom I\subseteq\cl\mathcal{C}$}.
\end{proposition}

In order to determine the subdifferentials of the function $z$, we
need at first to make $z$ explicit. To this aim, we set
$\ell:=\limsup_{s\uparrow\infty}(1/s)\ln e^{v(s)}p(s)$ and we stress
that $-\infty\le\ell\le z_o<\infty$ by assumption \ref{ass2}.  We also
recall that if $\mathcal{S}$ is an infinite set, then $f(s)/s$ has a
limit $r\in\Rl^d$ when $s$ goes to infinity through $\mathcal{S}$ by
assumption \ref{ass3}. When $\mathcal{S}$ is finite, then
$\ell=-\infty$ and we set $r:=f(s_o)/s_o$ for future convenience,
$s_o$ being an arbitrarily given element of $\mathcal{S}$.  Invoking
the Cauchy-Hadamard theorem we find that the series $\sum_{s\ge 1}
e^{k\cdot f(s)+v(s)-\zeta s}\,p(s)$ convergences if $\zeta>k\cdot
r+\ell$ and, in the case $\ell>-\infty$, divergences if $\zeta<k\cdot
r+\ell$. The properties of the function $z$ crucially depends on the
behavior of this series at $\zeta=k\cdot r+\ell$, so that it is
helpful to introduce the extended real number $\theta(k)$ defined by
\begin{eqnarray}
\nonumber
\theta(k):=\sum_{s\ge 1} e^{k\cdot f(s)+v(s)-(k\cdot r+\ell)s}\,p(s).
\end{eqnarray}
It is understood that $\theta(k)=\infty$ for all $k$ if
$\ell=-\infty$.  The function $\theta$ that maps each $k\in\Rl^d$ in
$\theta(k)$ is convex and lower semicontinuous because, in the case
$\ell>-\infty$, is the sum of finite positive convex functions. As a
consequence, the possible empty level set
\begin{eqnarray}
\nonumber
  \Theta:=\Big\{k\in\Rl^d\,:\,\theta(k)\le 1\Big\}
\end{eqnarray}
is convex and closed, and its complement $\Theta^c$ is open.
Necessary conditions for $\Theta$ to be nonempty are that
$\mathcal{S}$ is an infinite set and $\ell>-\infty$. It is a simple
exercise to verify that $\Theta=\Rl^d$ if and only if $\sum_{s\ge 1}
e^{v(s)-\ell s}\,p(s)\le 1$ and $f(s)=rs$ for all $s\in\mathcal{S}$.
The following lemma provides $z$ explicitly and is proved in
\ref{proof_zexplicit}.
\begin{lemma}
\label{zexplicit}
Pick $k\in\Rl^d$. Then
\begin{eqnarray}
\nonumber
  z(k)=
  \cases{
    k\cdot r+\ell & if $k\in\Theta$;\\
    \zeta & if $k\in\Theta^c$,
  }
\end{eqnarray}
where $\zeta>k\cdot r+\ell$ is the unique number that satisfies 
$\sum_{s\ge 1}e^{k\cdot f(s)+v(s)-\zeta s}\,p(s)=1$.
\end{lemma}

We are now ready to supply a complete description of the
subdifferentials of $z$, which is the task of the next proposition
whose proof is given in \ref{lemmaz}.  The proposition states that $z$
is analytic on the open set $\Theta^c$, so that $z$ is in particular
differentiable on $\Theta^c$. To connect subdifferentiability and
differentiability, we remind that $z$ is differentiable at a certain
point $k\in\Rl^d$ with gradient $\nabla z(k)$ if and only if $\partial
z(k)$ is a singleton containing only the vector $\nabla z(k)$ (see
\ref{convex}, proposition \ref{diff_der}). In order to understand the
content of the proposition, we also point out that the vector
\begin{equation}
\label{v(k)}
\nu(k):=\frac{\sum_{s\ge 1} f(s)\,e^{k\cdot f(s)+v(s)-z(k) s}\,p(s)}{\sum_{s\ge 1} s\,e^{k\cdot f(s)+v(s)-z(k) s}\,p(s)}
\end{equation}
exists whenever $\sum_{s\ge 1} s\,e^{k\cdot f(s)+v(s)-z(k)
  s}\,p(s)<\infty$ because $\|f(s)\|\le Ms$ for some constant
$M<\infty$ and all $s\in\mathcal{S}$ by assumption \ref{ass3}. This is
certainly the case if $k\in\Theta^c$ because $z(k)>k\cdot r+\ell$ when
$k\in\Theta^c$. From now on, we think of a vector $u\in\Rl^d$ as a
column vector and we denote by $u^{\rm T}$ its transpose.
\begin{proposition}
\label{lem:diffzeta}
The following conclusions hold:
\begin{enumerate}[{\upshape(a)}]
\setlength\itemsep{0.3em}
\item $z$ is analytic on $\Theta^c$ and $\nabla z(k)=\nu(k)$ for all
  $k\in\Theta^c$. The vector field $\nu$ that associates any
  $k\in\Theta^c$ with $\nu(k)$ is analytic and its Jacobian matrix
  $J(k)$ at $k$, which obviously is the Hessian matrix of $z$ at $k$,
  is given by
  \begin{eqnarray}
  \nonumber
J(k)=\frac{\sum_{s\ge 1} [f(s)-\nu(k)s][f(s)-\nu(k)s]^{\rm T}e^{k\cdot f(s)+v(s)-z(k) s}\,p(s)}{\sum_{s\ge 1} s\,e^{k\cdot f(s)+v(s)-z(k) s}\,p(s)};
\end{eqnarray}
\item if $k\in\Theta$ and $\theta(k)=1$, then $\partial z(k)=\{r\}$ or 
$\partial z(k)=\{(1-\alpha)r+\alpha\nu(k)\}_{\alpha\in[0,1]}$
according to the series $\sum_{s\ge 1} s\,e^{k\cdot f(s)+v(s)-z(k)
  s}\,p(s)$ diverges or converges;
\item if $k\in\Theta$ and $\theta(k)<1$, then $z$ is differentiable at $k$ and $\nabla z(k)=r$.
\end{enumerate}
\end{proposition}

To conclude this paragraph, we observe that while proposition
\ref{compute_Io} states that there exists at least one point $k$ with
the property that $w\in\partial z(k)$ when $w\in\ri(\dom I)$, nothing
is said about how many different points $k$ share this property. The
following lemma, which is proved in \ref{proof_lemma:unicity}, answers
the question.
\begin{lemma}
\label{lemma:unicity}
Let $w$ and $k$ be two points in $\Rl^d$ such that $w\in\partial
z(k)$. Let $h$ be another point in $\Rl^d$.  The following conclusions
hold:
\begin{enumerate}[{\upshape (a)}]
\item if $w\ne r$, then $w\in\partial z(h)$ if and only if $(h-k)\cdot[f(s)-rs]=0$ for all $s\in\mathcal{S}$;
\item if $w=r$ and $k\in\Theta^c$, then $\Theta=\emptyset$ and
  $w\in\partial z(h)$ if and only if
  $(h-k)\cdot[f(s)-rs]=0$ for all $s\in\mathcal{S}$;
\item if $w=r$ and $k\in\Theta$, then $w\in\partial z(h)$ if and only if $h\in\Theta$.
\end{enumerate}
\end{lemma}

\subsection{Analytical Properties of the Rate Function $I$}

Proposition \ref{compute_Io} offers a method to compute the rate
function $I$ that can count on the description of $\dom I$ given by
proposition \ref{domIo} and the expression of $\partial z(k)$ provided
for any $k$ by proposition \ref{lem:diffzeta}. In doing this,
proposition \ref{compute_Io} allows us to identify the main analytical
properties of $I$. Here we discuss these properties mostly assuming
that the set $\{f(s)/s\}_{s\in\mathcal{S}}$ is of full dimension,
i.e.\ its affine hull is $\Rl^d$. In this case, $\dom I$ is full
dimensional by proposition \ref{domIo}. Since the affine hull of any
subset of $\Rl^d$ is closed, the affine hull of
$\{f(s)/s\}_{s\in\mathcal{S}}$ contains $r$ and hence equals $\Rl^d$
if and only if there are exactly $d$ linearly independent vectors in
the set $\{f(s)-rs\}_{s\in\mathcal{S}}$.  This is the situation we
expect to face in common real applications and to which, however, we
can always reduce the problem\footnote{Assume that there are at most
  $d_o<d$ linearly independent vectors of the form
  $f(s_1)-rs_1,\ldots,f(s_{d_o})-rs_{d_o}$ with $s_1,\ldots,s_{d_o}$
  in $\mathcal{S}$. By expanding $f(s)-rs$ on these vectors, we can
  define a function $f_o:\{1,2,\ldots\}\cup\{\infty\}\to\Rl^{d_o}$ with
  the property that $f(s)-rs=Af_o(s)$ for all $s\in\mathcal{S}$, where
  $A\in\Rl^{d\times d_o}$ is a matrix whose $l$th column is
  $f(s_l)-rs_l$. By construction, the matrix $A$ has rank $d_o$, the
  values of $f_o$ on $\mathcal{S}$ are uniquely determined, and
  $f_o(s_1),\ldots,f_o(s_{d_o})$ is the canonical basis of
  $\Rl^{d_o}$. Moreover, $f(s_o)/s_o=0$ if $\mathcal{S}$ is finite or
  $f_o(s)/s$ has limit $0$ when $s$ goes to infinity through
  $\mathcal{S}$ if $\mathcal{S}$ is infinite.  It follows that $A^{\rm
    T}A\in\Rl^{d_o\times d_o}$ is invertible, where $A^{\rm T}$
  denotes the transpose of $A$, and that the affine hull of
  $\{f_o(s)/s\}_{s\in\mathcal{S}}$ is $\Rl^{d_o}$. Let $z$ and $I$ be
  the functions that (\ref{defzetafun}) and (\ref{defIorate})
  associate to $f$ and let $z_o$ and $I_o$ be the functions that
  (\ref{defzetafun}) and (\ref{defIorate}) associate to $f_o$. It is a
  simple exercise based solely on definitions (\ref{defzetafun}) and
  (\ref{defIorate}) to verify that $I(w)= I_o(A_ow+a_o)$ for all
  $w\in\dom I$, where $A_o:=(A^{\rm T}A)^{-1}A^{\rm T}$ and
  $a_o:=-A_or$. This way, the rate function $I$ can be computed
  starting from the rate function $I_o$ associated with a set
  $\{f_o(s)/s\}_{s\in\mathcal{S}}$ of full dimension.}.

If the set $\{f(s)/s\}_{s\in\mathcal{S}}$ is of full dimension and
$w\in\inter(\dom I)=\ri(\dom I)$, then there exists $k\in\Rl^d$ such
that $w\in\partial z(k)$ by part (c) of proposition
\ref{compute_Io}. If $w\ne r$, then such $k$ is unique by part (a) of
lemma \ref{lemma:unicity}. Thus, $\partial I(w)$ is a singleton by
part (b) of proposition \ref{compute_Io} and $I$ results
differentiable at $w$ (see \ref{convex}, proposition
\ref{diff_der}). This way, $I$ is differentiable on $\inter(\dom I)$
with the only possible exception of the point $r$, and hence it is
continuously differentiable by convexity (see \ref{convex},
proposition \ref{diff_cont}).  Still assuming that
$\{f(s)/s\}_{s\in\mathcal{S}}$ is full dimensional, if there exists
$k\in\Theta^c$ such that $r\in\partial z(k)$, then $I$ is
differentiable at $r$ because $I(r)=r\cdot k-z(k)+z(0)<\infty$ and
$\partial I(r)=\{k\}$ by part (b) of lemma \ref{lemma:unicity}. If
instead there is no point $k\in\Theta^c$ such that $r\in\partial
z(k)$, then $r\in\partial z(h)$ if and only if $h\in\Theta$ by
proposition \ref{lem:diffzeta}, so that $\partial I(r)=\Theta$ by part
(b) of proposition \ref{compute_Io}.  It is not excluded here that
$\Theta=\emptyset$. These arguments prove the following lemma.
\begin{lemma}
\label{lemma3}
Suppose $\{f(s)/s\}_{s\in\mathcal{S}}$, and hence {\upshape $\dom I$},
is of full dimension. Then, $I$ is continuously differentiable on
{\upshape $\inter(\dom I)$} possibly deprived of $r$. If $I$ is not
differentiable at $r$, then $\partial I(r)=\Theta$.
\end{lemma}
More can be said about the smoothness properties of $I$. Suppose
$\{f(s)/s\}_{s\in\mathcal{S}}$ is of full dimension.  Then, the
condition $f(s)=rs$ for all $s\in\mathcal{S}$ cannot hold true and
$\Theta\ne\Rl^d$ as a consequence. This way, $\Theta^c\ne\emptyset$
and the restriction of $\nu$ to $\Theta^c$ is an injective continuous
map by parts (a) and (b) of lemma \ref{lemma:unicity} since $\partial
z(k)=\{\nu(k)\}$ for every $k\in\Theta^c$. It follows by invariance of
domain (see \cite{Tammo}, theorem 10.3.7) that
\begin{eqnarray}
\nonumber
\mathcal{U}:=\Big\{\nu(k):k\in\Theta^c\Big\}
\end{eqnarray}
is a nonempty open subset of $\Rl^d$. Injectivity of $\nu$ also
entails the invertibility of the Jacobian matrix $J(k)$ for each
$k\in\Theta^c$ and the existence of an inverse map
$\mu:\mathcal{U}\to\Theta^c$ such that $\nu(\mu(w))=w$ for all
$w\in\mathcal{U}$.  The real analytic inverse function theorem (see
\cite{Krantz}, theorem 2.5.1) tells us the $\mu$ is analytic. For each
$w\in\mathcal{U}$ we have $w=\nu(\mu(w))=\nabla z(\mu(w))$ and part
(a) of proposition \ref{compute_Io} gives $I(w)=w\cdot
\mu(w)-z(\mu(w))+z(0)<\infty$, which shows that
$\mathcal{U}\subseteq\dom I$ and that $I$ is analytic on $\mathcal{U}$
thanks to analyticity of both $\mu$ and $z$. We have thus proved the
following result.
\begin{lemma}
\label{lemma_U2}
If $\{f(s)/s\}_{s\in\mathcal{S}}$ is of full dimension, then
$\mathcal{U}$ is a nonempty open subset of $\dom I$ on which $I$ is
analytic.
\end{lemma}

Now, besides the set $\{f(s)/s\}_{s\in\mathcal{S}}$ being of full
dimension, suppose that the function $z$ is differentiable
everywhere. In this case, part (c) of proposition \ref{compute_Io}
tells us that for any $w\in\inter(\dom I)$ there exists $k\in\Rl^d$
such that $\nabla z(k)=w$. Proposition \ref{lem:diffzeta} entails that
if $w\ne r$, then necessarily $k\in\Theta^c$ and $w=\nu(k)$, so that
$w\in\mathcal{U}$. It follows that $\mathcal{U}\subseteq\inter(\dom
I)$ differs from $\inter(\dom I)$ by at most the point $r$, which
implies that $I$ is analytic on $\inter(\dom I)$ possibly deprived of
$r$. If $r\notin\mathcal{U}$, then $r\in\partial z(k)$ if and only if
$k\in\Theta$ by proposition \ref{lem:diffzeta}, so that $\partial
I(r)=\Theta$ by part (b) of proposition \ref{compute_Io} and $r$ could
be a singular point of $I$. We point out that $I$ is strictly convex
on $\inter(\dom I)$ if $z$ is differentiable everywhere, even when
$\{f(s)/s\}_{s\in\mathcal{S}}$ is not of full dimension (see
\ref{convex}, proposition \ref{strict}).  This way, we have proved the
following theorem.
\begin{theorem}
\label{Iprop_z_diff}
Suppose $\{f(s)/s\}_{s\in\mathcal{S}}$ is of full dimension and $z$ is
differentiable throughout $\Rl^d$. Then, $I$ is strictly convex on
{\upshape $\inter(\dom I)$} and analytic on the open set
$\mathcal{U}$, which differs from {\upshape $\inter(\dom I)$} by at
most the point $r$. If $r\notin\mathcal{U}$, then $\partial
I(r)=\Theta$.
\end{theorem}

To conclude, let us investigate situations where $z$ is not
differentiable everywhere. If $z$ is not differentiable at some point,
then the set $\Theta$ is necessarily nonempty by proposition
\ref{lem:diffzeta}, so that $r\notin\mathcal{U}$ by part (b) of lemma
\ref{lemma:unicity} and $\partial I(r)=\Theta$. In this case, the rate
function $I$ cannot be strictly convex on $\ri(\dom I)$ and affine
stretches in its graph must emerge. In fact, if $k$ is a point where
$z$ is not differentiable, then $\partial z(k)$ must not be a
singleton and proposition \ref{lem:diffzeta} consequently tells us
that $k\in\Theta$, $\sum_{s\ge 1} s\,e^{k\cdot f(s)+v(s)-z(k)
  s}\,p(s)<\infty$, $\nu(k)\ne r$, and $\partial
z(k)=\{(1-\alpha)r+\alpha\nu(k)\}_{\alpha\in[0,1]}$.  This way, part
(a) of proposition \ref{compute_Io} stating that $I(w)=w\cdot
k-z(k)+z(0)$ for all $w\in\partial z(k)$ shows that $I$ maps affinely
the closed line segment in $\Rl^d $ from $r$ to $\nu(k)$ onto the
closed segment in $\Rl$ from $I(r)$ to $I(\nu(k))$. In spite of the
lack of strict convexity, $I$ is however continuously differentiable
on $\inter(\dom I)$ possibly deprived of $r$ and analytic on
$\mathcal{U}$, as stated by lemmas \ref{lemma3} and \ref{lemma_U2}.
We have thus demonstrated the following theorem.
\begin{theorem}
\label{Iprop_z_nondiff}
Suppose $\{f(s)/s\}_{s\in\mathcal{S}}$ is of full dimension and $z$ is
not differentiable on the whole $\Rl^d$. Then, $I$ is not strictly
convex on {\upshape $\inter(\dom I)$} but it is continuously
differentiable on {\upshape $\inter(\dom I)$} possibly deprived of
$r$, where $\partial I(r)=\Theta$. $I$ is analytic on $\mathcal{U}$
and $r\notin\mathcal{U}$.
\end{theorem}

\subsection{The Critical Constrained Pinning Model}
\label{sec:critical}

We have seen that non-differentiability of $z$ at some point causes an
affine stretch in the graph of $I$.  If this point is the origin, then
the large deviation principle stated by theorem \ref{mainth1} does not
completely describe the way the probabilities of the cumulative reward
$W_t$ decay. In fact, here we demonstrate that $W_t/t$ always
converges in probability to some constant vector $\rho\in\Rl^d$ but the
convergence is necessarily subexponential if $\partial z(0)$ is not a
singleton. Let us observe that the formula $I(w)=w\cdot k-z(k)+z(0)$
for $w\in\partial z(k)$ by part (a) of proposition \ref{compute_Io}
yields $I(w)=0$ when $w\in\partial z(0)$. The converse is also true
because if $I(w)=0$, then the bound $w\cdot k-z(k)+z(0)\le I(w)=0$
valid for every $k$ shows that $w$ is a subgradient of $z$ at the
origin.  These arguments give $I(w)=0$ if and only if $w\in\partial
z(0)$ and $I$ turns out to have more than one zero whenever $\partial
z(0)$ is not a singleton.

The probability that the scaled cumulative reward $W_t/t$ fluctuates
over closed sets that do not contain zeros of $I$ always decays
exponentially fast in $t$, as stated by the following lemma that is
proved in \ref{proof:exp_closed}.
\begin{lemma}
\label{lem:exp_closed}
Let $F\subset\Rl^d$ be a closed set disjoint from $\partial
z(0)$. Then, there exists a real number $\lambda>0$ such that
$\prob_t^c[W_t/t\in F]\le e^{-\lambda t}$ for all sufficiently large
$t$.
\end{lemma}
According to Ellis \cite{Ellisbook}, we say that $W_t/t$ {\it
  converges exponentially} to a constant vector $\rho$ if for any
$\delta>0$ there exists a real number $\lambda>0$ with the property
that for all sufficiently large $t$
\begin{equation}
  \prob_t^c\Big[\big\|W_t/t-\rho\big\|\ge\delta\Big]\le e^{-\lambda t}.
\label{expconv}
\end{equation}
Lemma \ref{lem:exp_closed} implies that $W_t/t$ converges
exponentially to a constant vector $\rho$ if $\partial z(0)$ is a
singleton containing $\rho$.  Proposition \ref{lem:diffzeta} tells us
that such $\rho$ equals $r$ if $\theta(0)\le 1$ and equals $\nu(0)$ if
$\theta(0)>1$, with $\nu(0)$ defined by (\ref{v(k)}).  The following
theorem, which is proved in \ref{propconvprob}, completes the picture
about the convergence in probability of $W_t/t$ towards an appropriate
constant vector $\rho$.
\begin{theorem}
\label{prop:conv}
Set $\rho:=r$ if $\theta(0)\le 1$ and $\partial z(0)$ is a singleton
and set $\rho:=\nu(0)$ if $\theta(0)>1$ or $\partial z(0)$ is not a
singleton. The following conclusions hold:
\begin{enumerate}[{\upshape(a)}]
\setlength\itemsep{0.3em}
\item $\lim_{t\uparrow\infty}\,\prob_t^c[\|W_t/t-\rho\|\ge \delta]=0$ for
  any $\delta>0$;
\item $W_t/t$ converges exponentially to $\rho$ if and only if $\partial
  z(0)$ is a singleton.
\end{enumerate}
\end{theorem}

Theorem \ref{prop:conv} tells us that the scaled cumulative reward
$W_t/t$ exhibits a complex behavior if $\partial z(0)$ is not a
singleton, whereby convergence in probability to $\rho$ is slower than
exponential. Lemma \ref{lem:exp_closed} teaches us that exponential
convergence is prevented by persistent fluctuations of $W_t/t$ over
the set $\partial z(0)$ where the rate function takes value zero.
Observing that $z(0)=\ell$ when $\theta(0):=\sum_{s\ge 1} e^{v(s)-\ell
  s}\,p(s)=1$ according to lemma \ref{zexplicit}, proposition
\ref{lem:diffzeta} states that necessary and almost sufficient
conditions for $\partial z(0)$ not to be a singleton are
\begin{equation}
\label{critical}
  \cases{
    \ell>-\infty; & \\
    \sum_{s\ge 1} e^{v(s)-\ell s}\,p(s)=1; & \\
    \sum_{s\ge 1} s\,e^{v(s)-\ell s}\,p(s)<\infty. &
   }
\end{equation}
We are led to call {\it critical} a constrained pinning model that
satisfies such conditions, which, we notice, do not involve the
function $f$ defining the deterministic rewards. Under the situation
identified by (\ref{critical}), $\partial z(0)$ is not a singleton,
and precisely is the closed line segment in $\Rl^d$ connecting $r$ to
$\nu(0)$ by part (b) of proposition \ref{lem:diffzeta}, whatever the
function $f$ is except for those peculiar $f$ satisfying $\nu(0)=r$.
According to the literature on large deviation principles in
Statistical Mechanics \cite{El,Phys1}, we call such line segment the
{\it phase transition segment}.  A critical constrained pinning model
is thus a constrained pinning model for which most scaled cumulative
rewards display persistent fluctuations, over their phase transition
segment, which lead to subexponential decays of probabilities that
cannot be captured by a large deviation principle.

\subsection{Large Fluctuations of $N_t$}
\label{application}

The theory developed in the last three paragraphs is well exemplified
in the case of the number $N_t:=\sum_{\tau=1}^tU_\tau$ counting the
renewals by $t$, which is the cumulative reward corresponding to the
deterministic rewards identified by the function $f(s)=1$ for all
$s\ge 1$. Large deviation principles for $N_t$ within non-constrained
renewal processes have been previously investigated by Glynn and Whitt
\cite{Glynn} under the regularity conditions of the G\"artner-Ellis
theorem and by Lefevere, Mariani, and Zambotti \cite{Mariani} in
general through a theory for the fluctuations of the empirical
measures of backward and forward recurrence times.  In this paragraph
we study the fluctuations of $N_t$ within the renewal models of
Statistical Mechanics described in Sect.\ \ref{sec:statmech}, which
are constrained pinning models where $p(s)>0$ and $v(s)=\beta$ for
every $s$, $\beta$ being a control parameter that can drive a phase
transition. Apart from the sign, this control parameter is the binding
energy $\epsilon$ in the Poland-Scheraga model, the chemical potential
$\mu$ in the Fisher-Felderhof model and in the Tokar-Dreyss\'e model,
and the entropic loss $\sigma$ in the Wako-Sait\^o-Mu\~noz-Eaton
model. We point out that assumption \ref{ass1} is satisfied because
$\mathcal{S}$ consists of all positive integers and assumption
\ref{ass3} is verified with $r=0$. According to the physical arguments
of Sect.\ \ref{sec:statmech}, we suppose that
$\limsup_{s\uparrow\infty}(1/s)\ln p(s)<\infty$, which guarantees that
also assumption \ref{ass2} is fulfilled.

Some preliminary considerations are in order.  It is convenient here
to make the dependence on $\beta$ explicit by writing $z_\beta$,
$\nu_\beta$, and $I_\beta$ in place of $z$, $\nu$, and $I$. According
to definition (\ref{defzetafun}), we have
$z_\beta(k):=\inf\{\zeta\in\Rl\,:\,\sum_{s\ge 1}e^{k+\beta-\zeta
  s}\,p(s)\le 1\}=z_0(k+\beta)$ for every $\beta$ and $k$. It follows
immediately from definition (\ref{defIorate}) that
$I_\beta(w)=I_0(w)-w\beta+z_0(\beta)-z_0(0)$ for all $\beta$ and $w$,
which is useful to disentangle $\beta$ and $w$. Setting
\begin{eqnarray}
\nonumber
  \beta_c:=\cases{
  -\infty & if $\ell=-\infty$ or $\ell>-\infty$ and $\sum_{s\ge 1}e^{-\ell s}\,p(s)=\infty$;\\
   -\ln\sum_{s\ge 1}e^{-\ell s}\,p(s) & if $\ell>-\infty$ and $\sum_{s\ge 1}e^{-\ell s}\,p(s)<\infty$
}
\end{eqnarray}
we find $\theta(k)=\sum_{s\ge 1} e^{k+\beta-\ell
  s}\,p(s)=e^{k+\beta-\beta_c}>1$ if $k>\beta_c-\beta$ and
$\theta(k)\le 1$ if $k\le\beta_c-\beta$ provided that
$\beta_c>-\infty$.  We notice that $\ell$ equals
$\limsup_{s\uparrow\infty}(1/s)\ln p(s)$ due to the finiteness of $v$.
Since $r=0$, lemma \ref{zexplicit} tells us that $z_\beta(k)>\ell$
satisfies the identity $\sum_{s\ge 1}e^{k+\beta-z_\beta(k) s}\,p(s)=1$
for any $k>\beta_c-\beta$ and that $z_\beta(k)=\ell$ for each
$k\le\beta_c-\beta$ when $\beta_c>-\infty$.  Proposition
\ref{lem:diffzeta} states that $z$ is differentiable at all points $k$
with the only possible exception of $k=\beta_c-\beta$ in the case
$\beta_c>-\infty$. Within this case, part (b) of proposition
\ref{lem:diffzeta} tells us that $z_\beta$ is differentiable at
$k=\beta_c-\beta$ when $\sum_{s\ge 1} s\,e^{-\ell s}\,p(s)=\infty$ and
is not differentiable at $k=\beta_c-\beta$ when $\sum_{s\ge 1}
s\,e^{-\ell s}\,p(s)<\infty$. In both situations we can write
$\partial z_\beta(\beta_c-\beta)=[0,w_c]$ with
\begin{eqnarray}
\nonumber
  w_c:=\cases{
  0 & if $\sum_{s\ge 1} s\,e^{-\ell s}\,p(s)=\infty$;\\
\frac{\sum_{s\ge 1}e^{-\ell s}\,p(s)}{\sum_{s\ge 1}s\,e^{-\ell s}\,p(s)} & if $\sum_{s\ge 1} s\,e^{-\ell s}\,p(s)<\infty$.
}
\end{eqnarray}
We stress that differentiability of $z_\beta$ and the value of $w_c$
are independent of $\beta$.

\subsubsection*{The phase transition}

Theorem \ref{prop:conv} shows that $N_t/t$ converges in probability to
a certain $\rho_\beta$ as $t$ is sent to infinity. This convergence is
exponential unless the model is critical, namely unless
$\beta_c>-\infty$, $\beta=\beta_c$, and $\sum_{s\ge 1}s\,e^{-\ell
  s}\,p(s)<\infty$.  The real number $\rho_\beta$ is given explicitly
by
\begin{eqnarray}
\nonumber
  \rho_\beta=\frac{\sum_{s\ge 1}e^{-z_\beta(0) s}\,p(s)}{\sum_{s\ge 1}s\,e^{-z_\beta(0) s}\,p(s)}
\end{eqnarray}
if $\beta_c=-\infty$ and by
\begin{eqnarray}
\nonumber
  \rho_\beta=\cases{
  0 & if $\beta<\beta_c$;\\
  w_c & if $\beta=\beta_c$;\\
  \frac{\sum_{s\ge 1}e^{-z_\beta(0) s}\,p(s)}{\sum_{s\ge 1}s\,e^{-z_\beta(0) s}\,p(s)} & if $\beta>\beta_c$.
}
\end{eqnarray}
when $\beta_c>-\infty$. We observe that the identity
$z_\beta(0)=z_0(\beta)$ valid for all $\beta$ in combination with part
(a) of proposition \ref{lem:diffzeta} tells us that $z_\beta(0)$ and
$\rho_\beta=z_0'(\beta)$ are analytic function of $\beta$ in the
region $\beta>\beta_c$, where $\rho_\beta$ is increasing with $\beta$
because $z_0$ is strictly convex. 

The way $\rho_\beta$ depends on $\beta$ reveals a phase transition
whenever $\beta_c>-\infty$, whereby the fraction of times that are
renewals changes from being zero to being positive at $\beta_c$ while
increasing $\beta$.  We have
$\lim_{\beta\downarrow\beta_c}\rho_\beta=w_c$, which shows that the
phase transition is continuous or discontinuous according to the
series $\sum_{s\ge 1}s\,e^{-\ell s}\,p(s)$ diverges or converges.  The
limit $\lim_{\beta\downarrow\beta_c}\rho_\beta=w_c$ is due to Abel's
theorem when $\sum_{s\ge 1}s\,e^{-\ell s}\,p(s)<\infty$. If instead
$\sum_{s\ge 1}s\,e^{-\ell s}\,p(s)=\infty$, then the fact that
$z_\beta(0)>\ell$ allows us to write for every $\beta>\beta_c$ and
$t\ge 1$ the bound
\begin{eqnarray}
\nonumber
  \rho_\beta=\frac{\sum_{s\ge 1}e^{-z_\beta(0) s}\,p(s)}{\sum_{s\ge 1}s\,e^{-z_\beta(0) s}\,p(s)}
\le\frac{\sum_{s\ge 1}e^{-\ell s}\,p(s)}{\sum_{s\ge 1}s\,e^{-z_\beta(0) s}\,p(s)}
  \le\frac{e^{-\beta_c}}{\sum_{s=1}^ts\,e^{-z_\beta(0) s}\,p(s)}.
\end{eqnarray}
If follows from here that $\lim_{\beta\downarrow\beta_c}\rho_\beta=0$
by sending $\beta$ to $\beta_c$ first and $t$ to infinity later.  We
point out that, according to our definition of critical constrained
pinning model, the model is critical at $\beta=\beta_c$ only if
undergoes a discontinuous phase transition.  In the case of the
Poland-Scheraga model with loop entropy $\sigma_l=al+b-c\ln l$ for all
$l\ge 1$ we recover well-known facts \cite{PolSch1}: there is a phase
transition corresponding to $\beta_c>-\infty$ only if $c>1$, and the
phase transition is continuous when $1<c\le 2$ and discontinuous when
$c>2$.

\subsubsection*{The rate function}

We describe now the rate function $I_\beta$. The convex hull
$\mathcal{C}$ of the full dimensional set $\{1/s\}_{s\in\mathcal{S}}$
is the half-open interval $(0,1]$ and proposition \ref{domIo} gives
  that $\dom I_\beta$ differs from the set $(0,1]$ for at most the
    point 0. In any case, we have $\inter(\dom I_\beta)=(0,1)$,
    $r=0\notin\inter(\dom I_\beta)$, and $I_\beta(w)=\infty$ for all
    $w\notin[0,1]$. Let us identify $\mathcal{U}$, which is an open
    subset of $(0,1)$ by lemma \ref{lemma_U2}. If $\beta_c=-\infty$ or
    $\beta_c>-\infty$ and $\sum_{s\ge 1} s\,e^{-\ell s}\,p(s)=\infty$,
    then $z_\beta$ is differentiable throughout $\Rl$ and theorem
    \ref{Iprop_z_diff} tells us that $\mathcal{U}=(0,1)$. In order to
    tackle the case $\beta_c>-\infty$ and $\sum_{s\ge 1} s\,e^{-\ell
      s}\,p(s)<\infty$, let us pick $w\in(0,1)$ and let us observe
    that there exists a point $k$ such that $w\in\partial z(k)$ by
    part (c) of proposition \ref{compute_Io}. Since $w\ne r=0$,
    proposition \ref{lem:diffzeta} tells us that $k\ge\beta_c-\beta$,
    in such a way that either $k>\beta_c-\beta$ and hence
    $w\in\mathcal{U}$ or $k=\beta_c-\beta$ and hence $w\in(0,w_c]$. In
      the second case $w\notin\mathcal{U}$ by the unicity of $k$
      stated by part (a) of lemma \ref{lemma:unicity}. These arguments
      show that necessarily $\mathcal{U}=(w_c,1)$.

Let us determine $I_\beta(w)$ when $w\in(0,1)$. If $w\in(0,1)$ but
$w\notin\mathcal{U}$, then we have necessarily $\beta_c>-\infty$,
$\sum_{s\ge 1} s\,e^{-\ell s}\,p(s)<\infty$, and $w\in(0,w_c]$, in
  such a way that $I_\beta(w)=w(\beta_c-\beta)-\ell+z_\beta(0)$ by
  part (a) of proposition \ref{compute_Io}. If instead
  $w\in\mathcal{U}$, then there exists $k>\beta_c-\beta$ with the
  property that $w=\nu_\beta(k)$ and
  $I_\beta(w)=wk-z_\beta(k)+z_\beta(0)$ follows again by part (a) of
  proposition \ref{compute_Io}. This formula can be
  simplified. Consider the function $\mathcal{V}$ that maps each
  $\zeta>\ell$ in
  \begin{eqnarray}
  \nonumber
\mathcal{V}(\zeta):=\frac{\sum_{s\ge 1} e^{-\zeta s}\,p(s)}{\sum_{s\ge 1} s\,e^{-\zeta s}\,p(s)}.
\end{eqnarray}
It is not difficult to verify that
$m^2\frac{\partial\mathcal{V}}{\partial\zeta}(\zeta)=\sum_{s\ge
  1}(s-m)^2q(s)>0$ for any $\zeta$ with $m:=\sum_{s\ge 1}s\,q(s)$ and
$q(s):=e^{-\zeta s}\,p(s)/\sum_{\sigma\ge 1}e^{-\zeta
  \sigma}\,p(\sigma)$ for all $s$, so that the function $\mathcal{V}$
is increasing. The condition $\nu_\beta(k)=w$ reads
$\mathcal{V}(z_\beta(k))=w$, so that stating that there exists
$k>\beta_c-\beta$ with the property that $\nu_\beta(k)=w$ is
tantamount to say that there exists a real number $\zeta>\ell$
independent of $\beta$ such that $\mathcal{V}(\zeta)=w$ and
$z_\beta(k)=\zeta$ for some $k>\beta_c-\beta$. The number $\zeta$ is
unique because $\mathcal{V}$ is increasing. This way, since
$e^{-k}=\sum_{s\ge 1}e^{\beta-z_\beta(k) s}\,p(s)$ for
$k>\beta_c-\beta$ by construction, we can express $I_\beta(w)$ in
terms of $\zeta$ as $I_\beta(w)=-w\ln\sum_{s\ge 1}e^{\beta-\zeta
  s}\,p(s)-\zeta+z_\beta(0)$.

In order to completely find out $I_\beta$, it remains to compute
$I_\beta(0)$ and $I_\beta(1)$. We demonstrate at first that
$\inf_{k\in\Rl}\{z_\beta(k)\}=\ell$, which gives the result
$I_\beta(0)=\sup_{k\in\Rl}\{-z_\beta(k)+z_\beta(0)\}=-\ell+z_\beta(0)$
and shows that $0\in\dom I_\beta$ if and only if $\ell>-\infty$.
Recall that $z_\beta(k)>\ell$ if $k>\beta_c-\beta$ and
$z_\beta(k)=\ell$ if $k\le\beta_c-\beta$ provided that
$\beta_c>-\infty$. Thus, $\inf_{k\in\Rl}\{z_\beta(k)\}=\ell$ is
trivial when $\beta_c>-\infty$. When $\beta_c=-\infty$, then we get
$\inf_{k\in\Rl}\{z_\beta(k)\}=\ell$ as a consequence of the limit
$\lim_{k\downarrow -\infty}z_\beta(k)=\ell$. In fact, $z_\beta$ is an
increasing function that satisfies $\sum_{s\ge 1}e^{k+\beta-z_\beta(k)
  s}\,p(s)=1$ for all $k\in\Rl$ when $\beta_c=-\infty$.  If it were
$\lim_{k\downarrow -\infty}z_\beta(k)=:\ell_o>\ell$, then we would
find $\lim_{k\downarrow -\infty}\sum_{s\ge 1}e^{-z_\beta(k)
  s}\,p(s)=\sum_{s\ge 1}e^{-\ell_os}\,p(s)<\infty$ by Abel's theorem,
thus obtaining $\lim_{k\downarrow -\infty}\sum_{s\ge
  1}e^{k+\beta-z_\beta(k) s}\,p(s)=0$ and contradicting the fact that
$\sum_{s\ge 1}e^{k+\beta-z_\beta(k) s}\,p(s)=1$ for every $k$.

As far as the value of $I(1)$ is concerned, part (d) of proposition
\ref{compute_Io} with any $u\in(0,1)$ gives
$I_\beta(1)=\lim_{w\uparrow 1}I_\beta(w)$. In order to compute this
limit, we observe that $\zeta$ solving the equation
$\mathcal{V}(\zeta)=w$ is an increasing function of $w\in\mathcal{U}$ that
goes to infinity when $w$ is sent to 1. In fact, $\mathcal{V}$ is an
increasing function that is bounded away from 1 on compact
intervals. As a consequence, $(1-w)\zeta=[1-\mathcal{V}(\zeta)]\zeta$
goes to 0 when $w$ is sent to 1 because for positive $\zeta>\zeta_o>\ell$
\begin{eqnarray}
\nonumber
  0\le \big[1-\mathcal{V}(\zeta)\big]\zeta&=&
  \frac{\sum_{s\ge 2} (s-1)\,e^{-\zeta (s-2)}\,p(s)}{\sum_{s\ge 1} s\,e^{-\zeta s}\,p(s)}\,\zeta e^{-2\zeta}\\
\nonumber
  &\le&\frac{\sum_{s\ge 2} (s-1)\,e^{-\zeta_o (s-2)}\,p(s)}{p(1)}\,\zeta e^{-\zeta}.
\end{eqnarray}
Then, writing $I_\beta(w)=-w\ln\sum_{s\ge 1}e^{\beta-\zeta
  (s-1)}\,p(s)-(1-w)\zeta+z_\beta(0)$ for every $w\in\mathcal{U}$ we realize that
$I_\beta(1)=\lim_{w\uparrow 1}I_\beta(w)=-\ln e^\beta p(1)+z_\beta(0)$.

In conclusion, by putting the pieces together, in the case
$\beta_c=-\infty$ we find
\begin{eqnarray}
\nonumber
  I_\beta(w)=\cases{
  -\ell+z_\beta(0) & if $w=0$;\\
  -w\ln\sum_{s\ge 1}e^{\beta-\zeta s}\,p(s)-\zeta+z_\beta(0) ~\mbox{ with }~\mathcal{V}(\zeta)=w & if $w\in(0,1)$;\\
  -\ln e^\beta p(1)+z_\beta(0) & if $w=1$;\\
  \infty & otherwise.
 }
\end{eqnarray}
Recalling that $w_c:=0$ if $\sum_{s\ge 1} s\,e^{-\ell
  s}\,p(s)=\infty$, in the case $\beta_c>-\infty$ we can write the
universal expression
\begin{eqnarray}
  \nonumber
  I_\beta(w)=\cases{
  w(\beta_c-\beta)-\ell+z_\beta(0) & if $w\in[0,w_c]$;\\
  -w\ln\sum_{s\ge 1}e^{\beta-\zeta s}\,p(s)-\zeta+z_\beta(0) ~\mbox{ with }~\mathcal{V}(\zeta)=w & if $w\in(w_c,1)$;\\
  -\ln e^\beta p(1)+z_\beta(0) & if $w=1$;\\
  \infty & otherwise.
}
\end{eqnarray}
By lemma \ref{lemma_U2}, the rate function $I_\beta$ is analytic on
$(0,1)$ with an affine stretch and a singularity at $w=w_c$ when
$\beta_c>-\infty$ and $\sum_{s\ge 1} s\,e^{-\ell s}\,p(s)<\infty$. It
is however continuously differentiable on $(0,1)$ by lemma
\ref{lemma3}.  The conditions $\beta_c>-\infty$, $\beta=\beta_c$, and
$\sum_{s\ge 1} s\,e^{-\ell s}\,p(s)<\infty$ that make critical the
model give $z_\beta(0)=\ell$ and $I_\beta(w)=0$ for all $w\in[0,w_c]$
as a consequence.

\appendix

\section{Proof of proposition \ref{King}}
\label{proof_Kingman}

For any positive integers $\tau$ and $\delta$ the variable
$U_{\tau+\delta}$ is independent of $U_1,\ldots,U_\tau$ and
distributed as $U_\delta$ conditional on the event that $\tau$ is a
renewal, namely conditional on $U_\tau=1$. This argument with
$\tau:=\tau_{m-1}$ and $\delta:=\tau_m-\tau_{m-1}$ yields
$\prob[U_{\tau_1}=\cdots=U_{\tau_m}=1]=\prob[U_{\tau_1}=\cdots=U_{\tau_{m-1}}=1]\cdot\prob[U_{\tau_m-\tau_{m-1}}=1]$,
which proves the proposition after iteration over $m$.  To see
formally that $U_{\tau+\delta}$ is independent of $U_1,\ldots,U_\tau$
and distributed as $U_\delta$ when $\tau$ is a renewal it suffices to
observe that if $\tau=T_n$ for some positive integer $n$, then
$T_i\le\tau$ for each $i\le n$ and $T_i>\tau$ for any $i>n$.  It
follows that $U_t$ with $t\le\tau$ takes the expression
$\sum_{i=1}^n\mathds{1}_{\{T_i=t\}}$ that depends only on
$S_1,\ldots,S_n$. At the same time, we find
$U_{\tau+\delta}=\sum_{i\ge
  n+1}\mathds{1}_{\{T_i=\tau+\delta\}}=\sum_{i\ge
  1}\mathds{1}_{\{S_{n+1}+\cdots+S_{n+i}=\delta\}}$, showing that
$U_{\tau+\delta}$ depends only on $S_{n+1},S_{n+2},\ldots$ through the
same formula that connects $U_\delta$ to $S_1,S_2,\ldots$.

\section{Convex Analysis Considerations in $\Rl^d$}
\label{convex}

This appendix lists the technical results from convex analysis that
are used in Sect.\ \ref{sec:I} and for which we refer to, e.g.,
\cite{Rockbook}.  Hereafter, let $C$ be a convex set in $\Rl^d$ and
let $\varphi$ be a proper convex function on $\Rl^d$.  The following
is a fundamental property of closures and relative interiors of convex
sets.
\begin{proposition}
\label{riC}
  Pick any {\upshape $x\in\cl C$} and {\upshape $y\in\ri C$}. Then,
{\upshape $\lambda x+(1-\lambda)y\in\ri C$} for all $\lambda\in[0,1)$.
\end{proposition}

The convex function $\varphi$ is continuous on the relative interior
$\ri(\dom\varphi)$ of its effective domain $\dom\varphi$, so that a
convex function finite on all of $\Rl^d$ is necessarily
continuous. The following result states certain continuity properties
of $\varphi$ on $\dom\varphi$.
\begin{proposition}
\label{riC_phi}
  Let $\varphi$ be a lower semicontinuous convex function and let $y$ be
a point in {\upshape $\ri(\dom\varphi)$}. Then, $\lim_{\lambda\uparrow
  1}\varphi(\lambda x+(1-\lambda)y)=\varphi(x)$ for all {\upshape
  $x\in\cl(\dom\varphi)$}.
\end{proposition}

A practical way to determine subgradients of convex functions relies
on directional derivatives. The one-sided directional derivative
$\varphi'(x;u)$ of $\varphi$ at $x$ with respect to a vector
$u\in\Rl^d$ is defined by
\begin{equation}
\label{dirder}
\varphi'(x;u):=\lim_{\epsilon\downarrow 0}\frac{\varphi(k+\epsilon u)-\varphi(x)}{\epsilon}.
\end{equation}
It exists as an extended real number and $\varphi(x+\epsilon u)\ge
\varphi(x)+\varphi'(x;u)\epsilon$ for all positive $\epsilon$ because
the difference quotient in (\ref{dirder}) is a non-decreasing function
of the parameter $\epsilon>0$ by convexity. The following result holds
true.
\begin{proposition}
\label{g_der}
  Let $x$ be a point where $\varphi$ is finite. Then, a vector $g$ is a
subgradient of $\varphi$ at $x$ if and only if $g\cdot u\le
\varphi'(x;u)$ for all $u\in\Rl^d$.
\end{proposition}
Subgradients and directional derivatives are related to
differentiability in the following way.
\begin{proposition}
\label{diff_der}
Let $x$ be a point where $\varphi$ is finite. Then, the following
three conditions are equivalent to each other:
\begin{enumerate}[{\upshape (a)}]
\item $\varphi$ is differentiable at $x$ with gradient $\nabla\varphi(x)$;
\item $\varphi$ has a unique subgradient $g$ at $x$;
\item there exists a vector $\nu$ (necessarily unique) such that
  $\varphi'(x;u)=\nu\cdot u$ for every $u\in\Rl^d$.
\end{enumerate}
If one of these conditions is fulfilled, then all these conditions are
satisfied and $\nabla\varphi(x)=g=\nu$.
\end{proposition}
According to the next result, the gradient mapping
$x\mapsto\nabla\varphi(x)$ is continuous on open sets where $\varphi$
is differentiable.
\begin{proposition}
\label{diff_cont}
Let $A\subseteq\Rl^d$ be an open set. If $\varphi$ is differentiable
at any point of $A$, then $\varphi$ is actually continuously
differentiable on $A$.
\end{proposition}

The Fenchel-Legendre transform, or conjugate, of $\varphi$ is the
lower semicontinuous proper convex function $\varphi^\star$ defined
for all $x^\star\in\Rl^d$ by
\begin{eqnarray}
\nonumber
\varphi^\star(x^\star):=\sup_{x\in\Rl^d}\big\{x^\star\cdot x-\varphi(x)\big\}.
\end{eqnarray}
Subdifferentials enter the theory of conjugate functions through the
following result.
\begin{proposition}
\label{sub_FL}
  The following two conditions are equivalent to each other for any
vectors $x$ and $x^\star$:
\begin{enumerate}[{\upshape (a)}]
\item $x^\star\in\partial\varphi(x)$;
\item $x^\star\cdot y-\varphi(y)$ achieves its supremum in $y$ at $y=x$.
\end{enumerate}
If $\varphi$ is lower semicontinuous, then one more condition can be
added to this list:
\begin{enumerate}
\item[{\upshape (c)}] $x\in\partial\varphi^\star(x^\star)$.
\end{enumerate}
\end{proposition}
Importantly, the subdifferentials of $\varphi$ are
very close to covering the effective domain of $\varphi^\star$ when
$\varphi$ is lower semicontinuous according to the following result.
\begin{proposition}
\label{domain_FL}
  If $\varphi$ is lower semicontinuous, then {\upshape
  $\ri(\dom\varphi^\star)\subseteq\cup_{x\in\Rl^d}\partial\varphi(x)\subseteq\dom\varphi^\star$}.
\end{proposition}
Differentiability of $\varphi$ everywhere is related to strict
convexity of $\varphi^\star$ by the following last result.
\begin{proposition}
\label{strict}
If $\varphi$ is finite everywhere, then $\varphi^\star$ is strictly
convex on $\ri(\dom\varphi^\star)$ if and only if $\varphi$ is
differentiable throughout $\Rl^d$.
\end{proposition}

\section{Proof of proposition \ref{domIo}}
\label{lemmadom}

Let $\mathcal{H}$ be the closed convex set in $\Rl^d$ defined by
\begin{eqnarray}
\nonumber
\mathcal{H}:=\Bigg\{w\in\Rl^d\,:\,k\cdot w\le\sup_{s\in\mathcal{S}}\bigg\{k\cdot \frac{f(s)}{s}\bigg\}
\mbox{ for all }k\in\Rl^d\Bigg\}.
\end{eqnarray}
We prove in the order that $\mathcal{C}\subseteq\dom I$, that
$\cl\mathcal{C}=\mathcal{H}$, and that $\dom I\subseteq\mathcal{H}$.

To begin with, let us observe that $\sum_{s\ge 1} e^{k\cdot
  f(s)+v(s)-\zeta s}\,p(s)$ is lower semicontinuous in $\zeta$ for any
given $k$ because is the sum of continuous positive functions. This
fact gives $\sum_{s\ge 1} e^{k\cdot f(s)+v(s)-z(k) s}\,p(s)\le 1$ by
definition (\ref{defzetafun}), which in turn implies $k\cdot
f(s)-z(k)s\le-v(s)-\ln p(s)$ for all $s\in\mathcal{S}$.  If $w$ is a
convex combination of the elements from
$\{f(s)/s\}_{s\in\mathcal{S}}$, then there exist integers
$s_1,\ldots,s_m$ in the support $\mathcal{S}$ of $p$ and positive real
numbers $\lambda_1,\ldots,\lambda_m$ such that
$w=\sum_{l=1}^m\lambda_lf(s_l)/s_l$ and $\sum_{l=1}^m\lambda_l=1$.
Let $M<\infty$ be a constant satisfying $-v(s_l)-\ln p(s_l)\le M s_l$
for every $l$ and pick an arbitrary point $k\in\Rl^d$.  Then,
$k\cdot f(s_l)-z(k)s_l\le -v(s_l)-\ln p(s_l)\le Ms_l$ for all $l$ and,
recalling that $\sum_{l=1}^m\lambda_l=1$ and $\lambda_l>0$ for each
$l$, we find
\begin{eqnarray}
\nonumber
k\cdot w-z(k) =\sum_{l=1}^m\lambda_l \bigg[k\cdot \frac{f(s_l)}{s_l}\bigg]-z(k)=
\sum_{l=1}^m\lambda_l \bigg[k\cdot \frac{f(s_l)}{s_l}-z(k)\bigg]\le M.
\end{eqnarray}
The arbitrariness of $k$ results in $I(w)\le M+z(0)<\infty$, showing
that $w\in\dom I$. The arbitrariness of $w$ implies
$\mathcal{C}\subseteq\dom I$.

If $w=\sum_{l=1}^m\lambda_l f(s_l)/s_l$ is as above with $\lambda_l>0$
for each $l$ and $\sum_{l=1}^m\lambda_l=1$, then for all $k\in\Rl^d$
we get
\begin{eqnarray}
\nonumber
k\cdot w=\sum_{l=1}^m\lambda_l\bigg[k\cdot \frac{f(s_l)}{s_l}\bigg]\le 
\sup_{s\in\mathcal{S}}\bigg\{k\cdot \frac{f(s)}{s}\bigg\}.
\end{eqnarray}
This gives $w\in\mathcal{H}$. Thus, $\mathcal{C}\subseteq\mathcal{H}$
is deduced from the arbitrariness of $w$ and
$\cl\mathcal{C}\subseteq\mathcal{H}$ follows since $\mathcal{H}$ is
closed. In order to show that $\cl\mathcal{C}=\mathcal{H}$ it remains
to prove that $\mathcal{H}\subseteq\cl\mathcal{C}$. By contradiction,
if there exists $u\in\mathcal{H}$ that is not contained in
$\cl\mathcal{C}$, then we can find a point $h$ and a number
$\epsilon>0$ such that $h\cdot w+\epsilon\le h\cdot u$ for all
$w\in\cl\mathcal{C}$ by the Hahn-Banach separation theorem.  In
particular, as $f(s)/s\in\cl\mathcal{C}$ for all $s\in\mathcal{S}$, we
obtain $h\cdot f(s)/s+\epsilon\le h\cdot u$ for each
$s\in\mathcal{S}$. This contradicts the fact that $k\cdot u\le
\sup_{s\in\mathcal{S}}\{k\cdot f(s)/s\}$ for every $k\in\Rl^d$ because
$u\in\mathcal{H}$.

To conclude, we prove that $\dom I\subseteq\mathcal{H}$.  Assume for a
moment to know that $z(k)\le \sup_{s\in\mathcal{S}}\{k\cdot
f(s)/s\}+|z_o|+\ln 2$ for all $k\in\Rl^d$ with $z_o$ given by
assumption \ref{ass2}. Then, definition (\ref{defIorate}) yields for
each point $k$ and positive real number $\eta$
\begin{eqnarray}
\nonumber
\eta k\cdot w\le I(w)+z(\eta k)-z(0)\le I(w)+\eta\,\sup_{s\in\mathcal{S}}\bigg\{k\cdot\frac{f(s)}{s}\bigg\}+|z_o|+\ln 2-z(0).
\end{eqnarray}
This way, if $I(w)<\infty$, then dividing by $\eta$ first and sending
$\eta$ to infinity later we obtain $k\cdot w\le
\sup_{s\in\mathcal{S}}\{k\cdot f(s)/s\}$ for any $k$. This shows that
$w\in\mathcal{H}$ whenever $w\in\dom I$. It remains to verify that
$z(k)\le \sup_{s\in\mathcal{S}}\{k\cdot f(s)/s\}+|z_o|+\ln 2$ for all
$k\in\Rl^d$. To this aim, fix a point $k$ in $\Rl^d$ and a real number
$\zeta<z(k)$, so that $\sum_{s\ge 1} e^{k\cdot f(s)+v(s)-\zeta
  s}\,p(s)>1$ by definition of $z(k)$. As $e^{v(s)}p(s)\le e^{z_os}$
for all $s$ by assumption \ref{ass2}, an integer $t\in\mathcal{S}$
exists with the property that $k\cdot f(t)-\zeta t>-z_o-\ln 2$. It
follows that $\zeta<k\cdot f(t)/t+(z_o+\ln
2)/t\le\sup_{s\in\mathcal{S}}\{k\cdot f(s)/s\}+|z_o|+\ln 2$, giving
$z(k)\le \sup_{s\in\mathcal{S}}\{k\cdot f(s)/s\}+|z_o|+\ln 2$ after
that $\zeta$ is sent to $z(k)$.

\section{Proof of lemma \ref{zexplicit}}
\label{proof_zexplicit}

Assume for a moment that $\Theta\ne\emptyset$ and pick
$k\in\Theta$. In this case $\ell>-\infty$ and we have $\sum_{s\ge 1}
e^{k\cdot f(s)+v(s)-\zeta s}\,p(s)=\infty$ or $\sum_{s\ge 1} e^{k\cdot
  f(s)+v(s)-\zeta s}\,p(s)\le \theta(k)\le 1$ according to
$\zeta<k\cdot r+\ell$ or $\zeta\ge k\cdot r+\ell$.  This way, a glance
at definition (\ref{defzetafun}) immediately tells us that
$z(k)=k\cdot r+\ell$.

Suppose now that $\Theta\ne\Rl^d$ and fix $k\in\Theta^c$.  The series
$\sum_{s\ge 1} e^{k\cdot f(s)+v(s)-\zeta s}\,p(s)$ defines a
non-increasing function in the variable $\zeta$. In the region
$\zeta>k\cdot r+\ell$, such function is finite, continuous, and
strictly decreasing to zero as $\zeta$ goes to infinity. Moreover, it
satisfies $\lim_{\zeta\downarrow k\cdot r+\ell}\sum_{s\ge 1} e^{k\cdot
  f(s)+v(s)-\zeta s}\,p(s)=\theta(k)>1$ by Abel's theorem. Then, we
deduce that there exists a unique real number $\zeta>k\cdot r+\ell$
solving the equation $\sum_{s\ge 1}e^{k\cdot f(s)+v(s)-\zeta
  s}\,p(s)=1$ and the value of $z$ at $k$ is exactly such number
$\zeta$ by definition (\ref{defzetafun}).

\section{Proof of proposition \ref{lem:diffzeta}}
\label{lemmaz}

\subsection*{Part {\upshape(a)}}

Let $G$ be the function that associates $k\in\Theta^c$ and
$\zeta>k\cdot r+\ell$ with the real number $G(k,\zeta):=\sum_{s\ge
  1}e^{k\cdot f(s)+v(s)-\zeta s}\,p(s)$. We prove that $G$ is
analytic. To this aim, fix $k\in\Theta^c$ and $\zeta>k\cdot r+\ell$
and set for brevity $c_0(s):=e^{k\cdot f(s)+v(s)-\zeta s}\,p(s)$ for
all $s\ge 1$.  Denote by $f_i$, $k_i$, $x_i$, and $\nu_i$ the $i$th
component respectively of $f$, $k$, a vector $x$ in $\Rl^d$, and
$\nu(k)$. As $\zeta>k\cdot r+\ell$, there exists $\delta_o>0$ such
that $\zeta\ge k\cdot r+\ell+(\|r\|+1)\delta_o$. It follows that if
$x\in\Rl^d$ and $y\in\Rl$ satisfy $\|x\|<\delta_o$ and $|y|<\delta_o$,
then $|x\cdot r|+|y|<(\|r\|+1)\delta_o$ and by the Cauchy's criterion
the series
\begin{eqnarray}
\nonumber
  \sum_{s\ge 1}\sum_{m_1\ge 0}\cdots\sum_{m_d\ge 0}\sum_{n\ge 0}\prod_{i=1}^d\frac{|x_if_i(s)|^{m_i}}{m_i!}\frac{|ys|^n}{n!}\,c_0(s)
  &=&\sum_{s\ge 1}e^{\sum_{i=1}^d|x_if_i(s)|+|ys|}\,c_0(s)\\
  \nonumber
  &=&\sum_{s\ge 1}e^{|x\cdot f(s)|+|ys|}\,c_0(s)
\end{eqnarray}
is convergent.  This way, Fubini's theorem allows us to freely
rearranged the order of summation to get
\begin{eqnarray}
\nonumber
G(k+x,\zeta+y)&=&\sum_{s\ge 1}e^{\sum_{i=1}^d x_if_i(s)-ys}\,c_0(s)\\
\nonumber
&=&\sum_{s\ge 1}\sum_{m_1\ge 0}\cdots\sum_{m_d\ge 0}\sum_{n\ge 0}\prod_{i=1}^d\frac{[x_if_i(s)]^{m_i}}{m_i!}\frac{(-ys)^n}{n!}\,c_0(s)\\
\nonumber
&=&\sum_{m_1\ge 0}\cdots\sum_{m_d\ge 0}\sum_{n\ge 0}\prod_{i=1}^d\frac{x_i^{m_i}}{m_i!}\frac{y^n}{n!}
\Bigg\{\sum_{s\ge 1}\prod_{i=1}^d[f_i(s)]^{m_i}(-s)^nc_0(s)\Bigg\}.
\end{eqnarray}
This formula shows that $G$ can be represented by a convergent power
series in an open neighborhood of the arbitrary given point
$(k,\zeta)$, thus demonstrating the analyticity of $G$. Moreover, it
gives
\begin{equation}
\label{111}
  \frac{\partial^{m_1+\cdots m_d+n}G}{\partial^{m_1}k_1\cdots\partial^{m_d}k_d\,\partial^n\zeta}(k,\zeta)=
  \sum_{s\ge 1}\prod_{i=1}^d[f_i(s)]^{m_i}(-s)^nc(s)
\end{equation}
for all non-negative integers $m_1,\ldots,m_d$ and $n$.

In particular, formula (\ref{111}) yields $\frac{\partial G}{\partial
  \zeta}(k,\zeta)=-\sum_{s\ge 1}s\,e^{k\cdot f(s)+v(s)-\zeta
  s}\,p(s)\ne 0$ for all $k\in\Theta^c$ and $\zeta>k\cdot
r+\ell$. This way, since $z(k)>k\cdot r+\ell$ and $G(k,z(k))=1$ for
each $k\in\Theta^c$, the real analytic implicit function theorem (see
\cite{Krantz}, theorem 2.3.5) tells us that $z$ is analytic on
$\Theta^c$.  By taking the derivative of $G(k,z(k))=1$ with respect to
$k_i$ we get for every index $i$ and point $k\in\Theta^c$
\begin{eqnarray}
\nonumber
  \frac{\partial z}{\partial k_i}(k)=-\frac{\frac{\partial G}{\partial k_i}(k,z(k))}{\frac{\partial G}{\partial\zeta}(k,z(k))}
  =\frac{\sum_{s\ge 1}f_i(s)\,e^{k\cdot f(s)+v(s)-z(k) s}\,p(s)}{\sum_{s\ge 1}s\,e^{k\cdot f(s)+v(s)-z(k) s}\,p(s)}=\nu_i(k).
\end{eqnarray}
The vector field $\nu$ that associates $k$ with $\nu(k)$ turns out to
be analytic on $\Theta^c$ inheriting this property from $z$.  As far
as the Jacobian matrix $J(k)$ of $\nu$ at $k$ is concerned, by taking
the derivative of $G(k,z(k))=1$ with respect to $k_i$ and $k_j$ we
find for every indices $i$ and $j$ and point $k\in\Theta^c$
\begin{eqnarray}
\nonumber
  \frac{\partial\nu_i}{\partial k_j}(k)&=&-\frac{\frac{\partial^2 G}{\partial k _i\partial k_j}(k,z(k))
    +\nu_j(k)\frac{\partial^2 G}{\partial k_i\partial \zeta}(k,z(k))
    +\nu_i(k)\frac{\partial^2 G}{\partial k_j\partial \zeta}(k,z(k))
    +\nu_i(k)\nu_j(k)\frac{\partial^2 G}{\partial^2 \zeta}(k,z(k))
  }{\frac{\partial G}{\partial
      \zeta}(k,z(k))}\\
\nonumber
  &=&\frac{\sum_{s\ge 1}[f_i(s)-\nu_i(k)s][f_j(s)-\nu_j(k)s]\,e^{k\cdot f(s)+v(s)-z(k) s}\,p(s)}{\sum_{s\ge
    1}s\,e^{k\cdot f(s)+v(s)-z(k) s}\,p(s)}.
\end{eqnarray}

\subsection*{Part {\upshape(b)}}

Assume that $\mathcal{S}$ is an infinite set and that $\ell>-\infty$,
otherwise $\Theta=\emptyset$ and there is nothing to prove, and pick
$k\in\Theta$. As discussed in \ref{convex}, a practical way to
determine the subgradients of the convex function $z$ at the point $k$
relies on the one-sided directional derivative $z'(k;u)$ with respect
to a vector $u\in\Rl^d$.  We have $z(k+\epsilon u)\ge
z(k)+z'(k;u)\epsilon$ for all positive $\epsilon$ and the vector $g$
is a subgradient of $z$ at $k$ if and only if $g\cdot u\le z'(k;u)$
for all $u\in\Rl^d$ (see \ref{convex}, proposition \ref{g_der}). The
function $z$ is differentiable at $k$ if and only if a vector $\nu$
(necessarily unique) exists so that $z'(k;u)=\nu\cdot u$ for every $u$
(see \ref{convex}, proposition \ref{diff_der}). If such a $\nu$
exists, then the gradient $\nabla z(k)$ of $z$ at $k$ is equal to
$\nu$.  Let us examine what happens when a vector $\nu\ne r$ exists so
that $z'(k;u)=\max\{r\cdot u, \nu\cdot u\}$ for all $u\in\Rl^d$. In
this case, $g:=(1-\alpha)r+\alpha \nu$ with any $\alpha\in[0,1]$
satisfies $g\cdot u\le z'(k;u)$ for each $u$, thus resulting in a
subgradient of $z$ at $k$.  Conversely, if $g$ is a subgradient of $z$
at $k$, then $g\cdot u\le\max\{r\cdot u, \nu\cdot u\}$ for every
$u$. It follows that $(g-r)\cdot u\le\max\{0, (\nu-r)\cdot u\}=0$ for
all vectors $u$ orthogonal to $\nu-r$, showing that a number $\alpha$
exists such that $g-r=\alpha(\nu-r)$, and hence $g=(1-\alpha)r+\alpha
\nu$. Taking $u=-(\nu-r)$ first and $u=\nu-r$ later in $g\cdot
u\le\max\{r\cdot u, \nu\cdot u\}$ we find that $\alpha\ge 0$ and that
$\alpha\le 1$, respectively. In conclusion, if $z(k;u)=\max\{r\cdot u,
\nu\cdot u\}$ for any $u\in\Rl^d$, then $g$ is a subgradient of $z$ at
$k$ if and only if there exists $\alpha\in[0,1]$ such that
$g=(1-\alpha)r+\alpha \nu$. This is true even in the case $\nu=r$, to
which a function $z$ differentiable at $k$ corresponds.  These
arguments tell us that in order to prove part (b) of the proposition
it suffices to check that if $\theta(k)=1$, then $z'(k;u)=r\cdot u$ or
$z'(k;u)=\max\{r\cdot u, \nu(k)\cdot u\}$ for any given $u\in\Rl^d$
depending on whether the series $\sum_{s\ge 1} s\,e^{k\cdot
  f(s)+v(s)-z(k) s}\,p(s)$ diverges or converges. Similarly, part (c)
follows if we prove that $z'(k;u)=r\cdot u$ for all $u$ when
$\theta(k)<1$.

Let us fix an arbitrary vector $u$ in $\Rl^d$. In order to simplify
next formulas, we set $c_\epsilon(s):=e^{(k+\epsilon u)\cdot
  f(s)+v(s)-z(k+\epsilon u) s}\,p(s)$ and
$\Delta_\epsilon(s):=\epsilon u\cdot f(s)-z(k+\epsilon u)s+z(k) s$ for
each $s\ge 1$ and $\epsilon\ge 0$. We notice that
$c_\epsilon(s)=e^{\Delta_\epsilon(s)}c_0(s)$ for all $s$ and
$\epsilon$ and that the equality $\nu(k)=\sum_{s\ge
  1}f(s)\,c_0(s)/\sum_{s\ge 1}s\,c_0(s)$ holds true according to
(\ref{v(k)}).  Given any $\epsilon$, lemma \ref{zexplicit} tells us
that $\sum_{s\ge 1}c_\epsilon(s)\le 1$ and that $\sum_{s\ge
  1}c_\epsilon(s)=1$ if $k+\epsilon u\in\Theta^c$. In addition, we
have $\sum_{s\ge 1}c_0(s)=\theta(k)$ because $z(k)=k\cdot r+\ell$ when
$k\in\Theta$. Finally, we observe that $\lim_{\epsilon\downarrow
  0}c_\epsilon(s)=c_0(s)$ for all $s$ thanks to the continuity of $z$.

Assuming that $\theta(k)=1$, so that $\sum_{s\ge
  1}c_0(s)=\theta(k)=1$, we now prove that $z'(k;u)=r\cdot u$ or
$z'(k;u)=\max\{r\cdot u, \nu(k)\cdot u\}$ depending on whether the
series $\sum_{s\ge 1} s\,e^{k\cdot f(s)+v(s)-z(k) s}\,p(s)$ diverges
or converges. The fact that $z(h)\ge h\cdot r+\ell$ for all
$h\in\Rl^d$ gives $z(k+\epsilon u)-z(k)\ge \epsilon r\cdot u$ for any
$\epsilon$, showing that $z'(k;u)\ge r\cdot u$.  Furthermore, the
bound $e^y\ge 1+y$ valid for every real number $y$ implies that if
$\sum_{s\ge 1} s\,c_0(s)<\infty$, then the series $\sum_{s\ge 1}
\Delta_\epsilon(s)\,c_0(s)$ exists for each $\epsilon>0$ and
\begin{eqnarray}
\nonumber
1&\ge&\sum_{s\ge 1} c_\epsilon(s)
=\sum_{s\ge 1} e^{\Delta_\epsilon(s)}\,c_0(s)\\
\nonumber
&\ge&\sum_{s\ge 1} c_0(s)+\sum_{s\ge 1} \Delta_\epsilon(s)\,c_0(s)=1+\sum_{s\ge 1} \Delta_\epsilon(s)\,c_0(s).
\end{eqnarray}
This yields $(1/\epsilon)\sum_{s\ge 1} \Delta_\epsilon(s)\,c_0(s)\le
0$ and sending $\epsilon$ to zero we obtain from here that $z'(k;u)\ge
\nu(k)\cdot u$ whenever $\sum_{s\ge 1} s\,c_0(s)<\infty$. These
arguments prove that $z'(k;u)\ge r\cdot u$ or $z'(k;u)\ge \max\{r\cdot
u, \nu(k)\cdot u\}$ according to $\sum_{s\ge 1} s\,c_0(s)=\infty$ or
$\sum_{s\ge 1} s\,c_0(s)<\infty$.

Let us deduce the opposite bounds $z'(k;u)\le r\cdot u$ if $\sum_{s\ge
  1} s\,c_0(s)=\infty$ and $z'(k;u)\le\max\{r\cdot u, \nu(k)\cdot u\}$
if $\sum_{s\ge 1} s\,c_0(s)<\infty$, which conclude the proof of part
(b) of the proposition.  Pick a number $\eta>0$ and observe that the
assumption that $f(s)/s$ has a limit $r\in\Rl^d$ when $s$ goes to
infinity through $\mathcal{S}$ ensures us that a positive integer
$\tau_o$ can be found with the property that $[f(s)-r s]\cdot u\le
\eta s$ for each $s\in\mathcal{S}$ larger than $\tau_o$.  Then, fix a
number $\epsilon>0$ and suppose for a moment that the condition
$k+\epsilon u\in\Theta^c$ is satisfied. Under this condition, $z$ is
differentiable at $k+\epsilon u$ as stated by part (a) of the present
proposition and it follows from convexity that for all $\tau\ge\tau_o$
\begin{eqnarray}
\nonumber
\frac{z(k+\epsilon u)-z(k)}{\epsilon}&\le&\nabla z(k+\epsilon u)\cdot u
=\frac{\sum_{s\ge 1} f(s)\cdot u \,c_\epsilon(s)}
{\sum_{s\ge 1} s\,c_\epsilon(s)}\\
\nonumber
&=&r\cdot u+\frac{\sum_{s\ge 1} [f(s)-r s]\cdot u \,c_\epsilon(s)}{\sum_{s\ge 1} s\,c_\epsilon(s)}\\
\nonumber
&\le&r\cdot u+\frac{\sum_{s=1}^\tau [f(s)-r s]\cdot u \,c_\epsilon(s)}{\sum_{s\ge 1} s\,c_\epsilon(s)}+\eta.
\end{eqnarray}
We get from here that for any integer $t\ge 1$
\begin{eqnarray}
\nonumber
\frac{z(k+\epsilon u)-z(k)}{\epsilon}
&\le&r\cdot u+\frac{\max\big\{0,\sum_{s=1}^\tau [f(s)-r s]\cdot u \,c_\epsilon(s)\big\}}
{\sum_{s\ge 1} s \,c_\epsilon(s)}+\eta\\
\nonumber
&\le&r\cdot u+\frac{\max\big\{0,\sum_{s=1}^\tau [f(s)-r s]\cdot u \,c_\epsilon(s)\big\}}
{\sum_{s=1}^t s\,c_\epsilon(s)}+\eta.
\end{eqnarray}
This inequality holds even if $k+\epsilon u\in\Theta$, and hence
whatever $k+\epsilon u$ is, since $z(k+\epsilon u)-z(k)=\epsilon
r\cdot u$ when $k+\epsilon u\in\Theta$. This way, we can send
$\epsilon$ to zero finding that for all $\tau\ge\tau_o$ and $t\ge 1$
\begin{eqnarray}
\nonumber
z'(k;u)\le r\cdot u+\frac{\max\big\{0,\sum_{s=1}^\tau [f(s)-r s]\cdot u\,c_0(s)\big\}}
{\sum_{s=1}^t s\,c_0(s)}+\eta.
\end{eqnarray}
Recall that $\lim_{\epsilon\downarrow 0}c_\epsilon(s)=c_0(s)$ for each
$s$.  At this point, sending first $t$ to infinity, then $\tau$ to
infinity, and finally $\eta$ to zero, we get $z'(k;u)\le r\cdot u$ or
$z'(k;u)\le\max\{r\cdot u, \nu(k)\cdot u\}$ according to the series
$\sum_{s\ge 1} s\,c_0(s)$ diverges or converges.

\subsection*{Part {\upshape(c)}}

Supposing that $\theta(k)<1$, so that $\sum_{s\ge
  1}c_0(s)=\theta(k)<1$, here we show that $z'(k;u)=r\cdot u$. If
there exists a real number $\epsilon_o>0$ with the property that
$k+\epsilon_o u\in\Theta$, then $k+\epsilon u\in\Theta$ for every
$\epsilon\in(0,\epsilon_o)$ since $k+\epsilon
u=(1-\epsilon/\epsilon_o)k+(\epsilon/\epsilon_o)(k+\epsilon_o u)$ and
both $k$ and $k+\epsilon_o u$ belong to the convex set
$\Theta$. Consequently, $z(k+\epsilon u)=(k+\epsilon u)\cdot r+\ell$
for all $\epsilon\in(0,\epsilon_o)$ and $z'(k;u)=r\cdot u$ follows
immediately. The non trivial case is when the number $\epsilon_o$ does
not exist, namely when $k+\epsilon u\in\Theta^c$ for all
$\epsilon>0$. However, in such case there exists a subsequence
$\{s_i\}_{i\ge 1}$ of $\mathcal{S}$ diverging to infinity with the
property that $z'(k;u)s_i<u\cdot f(s_i)$ for any $i$ as we shall prove
in a moment.  This fact yields
\begin{eqnarray}
\nonumber
z'(k;u)\le \lim_{i\uparrow\infty}\frac{u\cdot f(s_i)}{s_i}=r\cdot u.
\end{eqnarray}
On the other hand, since $z(h)\ge h\cdot r+\ell$ for all $h\in\Rl^d$
and $z(k)=k\cdot r+\ell$, we also have $z'(k;u)\ge r\cdot u$. This
way, $z'(k;u)=r\cdot u$ even when the above $\epsilon_o$ does not
exist and the proof of part (c) of the proposition is concluded.

Assume that $k+\epsilon u\in\Theta^c$ for all $\epsilon>0$.  We prove
that there exists a subsequence $\{s_i\}_{i\ge 1}$ of $\mathcal{S}$
diverging to infinity such that $z'(k;u)s_i<u\cdot f(s_i)$ for any $i$
by contradiction. Once again, we set $c_\epsilon(s):=e^{(k+\epsilon
  u)\cdot f(s)+v(s)-z(k+\epsilon u) s}\,p(s)$ and
$\Delta_\epsilon(s):=\epsilon u\cdot f(s)-z(k+\epsilon u)s+z(k) s$ for
each $s\ge 1$ and $\epsilon\ge 0$. If an integer $t\ge 1$ with the
property that $z'(k;u)s\ge u\cdot f(s)$ for all $s\in\mathcal{S}$
larger than $t$ exists, then $z(k+\epsilon u)-z(k)\ge
z'(k;u)\epsilon\ge \epsilon u\cdot f(s)/s$ for all those $s$ and
$\epsilon>0$ by convexity.  This means that $\Delta_\epsilon(s)\le 0$
for any $s\in\mathcal{S}$ larger than $t$ and $\epsilon>0$.  On the
other hand, we have $\sum_{s\ge 1} c_\epsilon(s)=1$ for every
$\epsilon>0$ as $k+\epsilon u\in\Theta^c$ by hypothesis. It follows
that $1=\sum_{s\ge 1} c_\epsilon(s)\le\sum_{s=1}^t
e^{\Delta_\epsilon(s)}c_0(s)+\sum_{s=t+1}^\infty c_0(s)$ for every
$\epsilon>0$.  This way, sending $\epsilon$ to zero we get $\sum_{s\ge
  1} c_0(s)\ge 1$, which contradicts the fact that $\sum_{s\ge 1}
c_0(s)<1$.

\section{Proof of lemma \ref{lemma:unicity}}
\label{proof_lemma:unicity}

\subsection*{Part {\upshape(a)}}

Fix $w\ne r$ and $k\in\Rl^d$ in such a way that $w\in\partial
z(k)$. By proposition \ref{lem:diffzeta}, the fact that $w\ne r$
implies $\theta(k)\ge 1$ and $\sum_{s\ge 1}s\,e^{k\cdot f(s)+v(s)-z(k)
  s}\,p(s)<\infty$, so that $\nu(k)$ is well defined.  Let $h$ be an
arbitrary point in $\Rl^d$.  If the condition $(h-k)\cdot[f(s)-rs]=0$
for every $s\in\mathcal{S}$ is satisfied, then $\partial z(h)=\partial
z(k)$ and hence $w\in\partial z(h)$.  Indeed, it is a simple exercise
to verify that such condition entails $\theta(h)=\theta(k)\ge 1$,
$z(h)=z(k)+(h-k)\cdot r$ based on definition (\ref{defzetafun}),
$\sum_{s\ge 1}s\,e^{h\cdot f(s)+v(s)-z(h) s}\,p(s)=\sum_{s\ge
  1}s\,e^{k\cdot f(s)+v(s)-z(k) s}\,p(s)<\infty$, and
$\nu(h)=\nu(k)$. In particular, the results $\theta(h)\ge 1$,
$\sum_{s\ge 1}s\,e^{k\cdot f(s)+v(s)-z(k) s}\,p(s)<\infty$, and
$\nu(h)=\nu(k)$ combined with proposition \ref{lem:diffzeta}
necessarily yield $\partial z(h)=\partial z(k)$.

Assume now that $w\in\partial z(h)$. We prove that
$(h-k)\cdot[f(s)-rs]=0$ for every $s\in\mathcal{S}$ as a consequence.
We have $z(x)\ge z(h)+w\cdot(x-h)$ for all $x\in\Rl^d$ since $w$ is a
subgradient of $z$ at $h$ and, similarly, $z(x)\ge
z(k)+w\cdot(x-k)$. It follows in particular that
$z(h)-z(k)=w\cdot(h-k)$ and that $z(\lambda h+(1-\lambda)k)\ge
z(k)+\lambda w\cdot(h-k)$ for all $\lambda\in\Rl$.  These two
relations give $z(\lambda h+(1-\lambda)k)\ge \lambda
z(h)+(1-\lambda)z(k)$, which combined with convexity entails
$z(\lambda h+(1-\lambda)k)=\lambda z(h)+(1-\lambda)z(k)$ for every
$\lambda\in[0,1]$. Suppose for a moment that $k\in\Theta^c$ and recall
that $z$ is analytic on $\Theta^c$. Then, $\lambda
h+(1-\lambda)k\in\Theta^c$ for all sufficiently small $\lambda$ as
$\Theta^c$ is open and by taking the second derivative of $z(\lambda
h+(1-\lambda)k)=\lambda z(h)+(1-\lambda)z(k)$ with respect to
$\lambda$ and sending $\lambda$ to zero we find $(h-k)\cdot
J(k)(h-k)=0$, $J(k)$ being the Hessian matrix of $z$ at $k$. By part
(a) of proposition \ref{lem:diffzeta}, the condition $(h-k)\cdot
J(k)(h-k)=0$ is tantamount to
\begin{eqnarray}
\nonumber
\sum_{s\ge 1} \Big\{(h-k)\cdot[f(s)-\nu(k)s]\Big\}^2e^{k\cdot f(s)+v(s)-z(k) s}\,p(s)=0.
\end{eqnarray}
It follows from here that $(h-k)\cdot[f(s)-\nu(k)s]=0$ whenever
$s\in\mathcal{S}$, which gives $(h-k)\cdot[f(s)/s-f(\sigma)/\sigma]=0$
for each $s$ and $\sigma$ in $\mathcal{S}$. By setting $\sigma$ equal
to $s_o$ if $\ell=-\infty$ or by sending $\sigma$ to infinity if
$\ell>-\infty$, we get $(h-k)\cdot[f(s)-rs]=0$ for all
$s\in\mathcal{S}$. The same conclusion is achieved by changing $k$
with $h$ if $h\in\Theta^c$.

It remains to tackle the case where both $h$ and $k$ belong to
$\Theta$. In this case, $\ell>-\infty$ because
$\Theta\ne\emptyset$. Moreover, $z(h)=h\cdot r+\ell$ and $z(k)=k\cdot
r+\ell$, so that the above condition $z(h)-z(k)=w\cdot(h-k)$ becomes
$(w-r)\cdot(h-k)=0$. Since $h\in\Theta$, $w\in\partial z(h)$, and
$w\ne r$, proposition \ref{lem:diffzeta} tells us that necessarily
there exists $\alpha>0$ such that
$w=(1-\alpha)r+\alpha\nu(h)$. Similarly, there exists $\beta>0$ such
that $w=(1-\beta)r+\beta\nu(k)$. These two identities combined with
$(w-r)\cdot(h-k)=0$ yield
$[\nu(h)-r]\cdot(h-k)=[\nu(k)-r]\cdot(h-k)=0$.  Write
$\bar{f}(s):=f(s)-rs$ for all $s$, $\bar{\nu}(k):=\nu(k)-r$, and
$\bar{\nu}(h):=\nu(h)-r$ for brevity.  Let $\mathcal{S}_+$ and
$\mathcal{S}_-$ be the subsets of $\mathcal{S}$ where
$(h-k)\cdot\bar{f}(s)\ge 0$ and $(h-k)\cdot\bar{f}(s)\le 0$,
respectively. Using first the fact that $(h-k)\cdot\bar{\nu}(h)=0$,
namely $\sum_{s\ge 1}(h-k)\cdot\bar{f}(s)\,e^{h\cdot
  \bar{f}(s)+v(s)-\ell s}\,p(s)=0$, and later the fact that
$(h-k)\cdot\bar{\nu}(k)=0$, namely $\sum_{s\ge
  1}(h-k)\cdot\bar{f}(s)\,e^{k\cdot \bar{f}(s)+v(s)-\ell s}\,p(s)=0$,
we get
\begin{eqnarray}
\nonumber
\sum_{s\in\mathcal{S}_+}(h-k)\cdot\bar{f}(s)\,e^{h\cdot \bar{f}(s)+v(s)-\bar{z}(h) s}\,p(s)&=&
-\sum_{s\in\mathcal{S}_-}(h-k)\cdot\bar{f}(s)\,e^{h\cdot \bar{f}(s)+v(s)-\bar{z}(h) s}\,p(s)\\
\nonumber
&=&-\sum_{s\in\mathcal{S}_-}(h-k)\cdot\bar{f}(s)\,e^{(h-k)\cdot \bar{f}(s)+k\cdot \bar{f}(s)+v(s)-\bar{z}(k) s}\,p(s)\\
\nonumber
&\le&-\sum_{s\in\mathcal{S}_-}(h-k)\cdot\bar{f}(s)\,e^{k\cdot \bar{f}(s)+v(s)-\bar{z}(k) s}\,p(s)\\
\nonumber
&=&\sum_{s\in\mathcal{S}_+}(h-k)\cdot\bar{f}(s)\,e^{k\cdot \bar{f}(s)+v(s)-\bar{z}(k) s}\,p(s).
\end{eqnarray}
This bound can be recast as
\begin{eqnarray}
\nonumber
\sum_{s\in\mathcal{S}_+}(h-k)\cdot\bar{f}(s)\,\Big[e^{(h-k)\cdot \bar{f}(s)}-1\Big]\,e^{k\cdot \bar{f}(s)+v(s)-\bar{z}(k) s}\,p(s)\le 0,
\end{eqnarray}
which shows that $(h-k)\cdot\bar{f}(s)=0$ for each
$s\in\mathcal{S}_+$. A similar argument gives $(h-k)\cdot\bar{f}(s)=0$
for every $s\in\mathcal{S}_-$, so that
$(h-k)\cdot\bar{f}(s)=(h-k)\cdot[f(s)-rs]=0$ for all
$s\in\mathcal{S}$.

\subsection*{Part {\upshape(b)}}

Suppose that there exists $k\in\Theta^c$ such that $r\in\partial
z(k)=\{\nu(k)\}$ and bear in mind that $z(k)-k\cdot r>\ell$ and
$\sum_{s\ge 1}e^{k\cdot f(s)+v(s)-z(k) s}\,p(s)=1$ for such $k$ by
lemma \ref{zexplicit}. The condition $\nu(k)=r$ is tantamount to
$\sum_{s\ge 1}\bar{f}(s)\,e^{k\cdot f(s)+v(s)-z(k) s}\,p(s)=0$ with
$\bar{f}(s):=f(s)-rs$ for all $s$. This way, making use of the bound
$z(k)-k\cdot r>\ell$ first and on the bound $e^y\ge 1+y$ valid for all
$y\in\Rl$ later, we find for each $h\in\Rl^d$
\begin{eqnarray}
\nonumber
\theta(h)&=&\sum_{s\ge 1}e^{h\cdot \bar{f}(s)+v(s)-\ell s}\,p(s)>\sum_{s\ge 1}e^{(h-k)\cdot \bar{f}(s)}\,e^{k\cdot f(s)+v(s)-z(k) s}\,p(s)\\
  \nonumber
  &\ge&\sum_{s\ge 1}e^{k\cdot f(s)+v(s)-z(k) s}\,p(s)
  +(h-k)\cdot \sum_{s\ge 1}\bar{f}(s)\,e^{k\cdot f(s)+v(s)-z(k) s}\,p(s)=1.
\end{eqnarray}
It follows from here that $\Theta=\emptyset$.

Let $h$ be a point in $\Rl^d=\Theta^c$. If the condition
$(h-k)\cdot[f(s)-rs]=0$ for every $s\in\mathcal{S}$ is satisfied, then
it is immediate to verify that $\nu(h)=\nu(k)=r$. If instead
$r=\nu(h)$, then $z(\lambda h+(1-\lambda)k)=\lambda
z(h)+(1-\lambda)z(k)$ for all $\lambda\in[0,1]$ as before. This way,
by taking the second derivative with respect to $\lambda$ and by
repeating the previous arguments, we find $(h-k)\cdot[f(s)-rs]=0$ for
all $s\in\mathcal{S}$.

\subsection*{Part {\upshape(c)}}

Suppose that $w=r$ and and that $k\in\Theta$. The latter in particular
means that $\Theta\ne\emptyset$. Pick $h\in\Rl^d$. If $h\in\Theta$,
then $r\in\partial z(h)$ by part (b) and (c) of proposition
\ref{lem:diffzeta}. If $r\in\partial z(h)$ and $h\in\Theta^c$, then
$\Theta=\emptyset$ by part (b), which is a contradiction. This way,
$r\in\partial z(h)$ implies $h\in\Theta$.

\section{Proof of lemma \ref{lem:exp_closed}}
\label{proof:exp_closed}

The lemma is due to part (c) of theorem \ref{mainth1} and the fact
that $\inf_{w\in F}\{I(w)\}>0$ since $F\cap\partial
z(0)=\emptyset$. The latter is obvious if $I(w)\ge 1$ for all $w\in
F$. If instead $I(w)<1$ for some $w\in F$, then the nonempty set
$K:=\{w\in F:I(w)\le 1\}$ is compact because $I$ is a good rate
function and, as a consequence, $I$ attains a minimum over $K$ by
lower semicontinuity. This means that there exists $u\in K$ such that
$I(w)\ge I(u)$ for all $w\in K$, and $I(w)\ge I(u)$ for all $w\in F$
follows as $I(u)\le 1$. We find $\inf_{w\in F}\{I(w)\}=I(u)>0$ because
$u\notin\partial z(0)$ when $u\in K$.

\section{Proof of theorem \ref{prop:conv}}
\label{propconvprob}

\subsection*{Part {\upshape(b)}}

We prove part (b) first. We already know from lemma
\ref{lem:exp_closed} that if $\partial z(0)$ is a singleton, then
$W_t/t$ converges exponentially to $\rho:=r$ when $\theta(0)\le 1$ and
to $\rho:=\nu(0)$ when $\theta(0)>1$.  Conversely, if (\ref{expconv})
holds for a fixed $\delta>0$ and the corresponding $\lambda>0$, then
part (b) of theorem \ref{mainth1} shows that $-I(w)\le
\liminf_{t\uparrow\infty}(1/t)\ln\prob_t^c[\|W_t/t-\rho\|>\delta]\le-\lambda$
whenever $\|w-\rho\|>\delta$. This implies that if $w\in\partial
z(0)$, so that $I(w)=0$, then $\|w-\rho\|\le\delta$ and the
arbitrariness of $\delta$ gives $w=\rho$.

\subsection*{Part {\upshape(a)}}

Part (a) follows from part (b) when $\partial z(0)$ is a singleton, so
that it remains to verity part (a) when $z$ is not differentiable at
the origin. Set $p_o(s):=e^{v(s)-\ell s}\,p(s)$ for all $s\ge
1$. Proposition \ref{lem:diffzeta} states that necessary conditions
for $z$ not to be differentiable at the origin are $\ell>-\infty$,
$\theta(0)=\sum_{s\ge 1}p_o(s)=1$, and $\sum_{s\ge 1}s\,p_o(s)<\infty$
as $z(0)=\ell$ by lemma \ref{zexplicit} when $\theta(0)=1$.  Let us
consider for a moment a new probability space
$(\Omega_o,\mathcal{F}_o,\prob_o)$ where a sequence $\{S_i\}_{i\ge 1}$
of independent waiting times distributed according to the new
distribution $p_o$ is given. Denoting by $\Ex_o$ the expectation under
$\prob_o$, we have $\Ex_o[S_1]=\sum_{s\ge 1} s\,p_o(s)<\infty$ and
$\Ex_o[\|f(S_1)\|]<\infty$ since $\|f(S_1)\|\le MS_1$ with some
positive constant $M<\infty$ and full probability by assumption
\ref{ass3}. We observe that $\Ex_o[f(S_1)]/\Ex_o[S_1]=\nu(0)$.  The
probability space $(\Omega_o,\mathcal{F}_o,\prob_o)$ fulfills the
following important properties:
$\lim_{t\uparrow\infty}\Ex_o[U_t]=1/\Ex_o[S_1]$ and
\begin{equation}
\lim_{t\uparrow\infty}\,\prob_o\Bigg[\bigg\|\frac{\sum_{i=1}^{N_t}f(S_i)}{\sum_{i=1}^{N_t}S_i}-\nu(0)\bigg\|\ge\delta \Bigg]=0
\label{libconv}
\end{equation}
for any $\delta>0$, $N_t$ being the number of renewals by $t$.  Since
$\sum_{s\ge 1} p_o(s)=1$, the limit
$\lim_{t\uparrow\infty}\Ex_o[U_t]=1/\Ex_o[S_1]$ is established by
applying the renewal theorem (see \cite{Feller1}, theorem 1 in Chapter
XIII.10) to the renewal equation
$\Ex_o[U_t]=\sum_{s=1}^tp_o(s)\,\Ex_o[U_{t-s}]$ valid for every $t\ge
1$. This equation is deduced by conditioning on $T_1=S_1$ and then by
using the fact that a renewal process starts over at every renewal.
The limit (\ref{libconv}) is due to the strong law of large
numbers. In fact, the strong law of large numbers tells us that
$\lim_{n\uparrow\infty}(1/n)\sum_{i=1}^n S_i=\Ex_o[S_1]$ and
$\lim_{n\uparrow\infty}(1/n)\sum_{i=1}^n f(S_i)=\Ex_o[f(S_1)]$
$\prob_o$-almost surely since $\Ex_o[S_1]<\infty$ and
$\Ex_o[\|f(S_1)\|]<\infty$. On the other hand,
$\lim_{t\uparrow\infty}N_t=\infty$ $\prob_o$-almost surely because the
event where one of the waiting times is infinite has probability zero
with respect to the probability measure $\prob_o$. This way, we get
$\lim_{t\uparrow\infty}\sum_{i=1}^{N_t} f(S_i)/\sum_{i=1}^{N_t}
S_i=\Ex_o[f(S_1)]/\Ex_o[S_1]=\nu(0)$ $\prob_o$-almost surely and
(\ref{libconv}) follows from the fact that almost sure convergence
implies converge in probability.

The features of the probability space
$(\Omega_o,\mathcal{F}_o,\prob_o)$ allow us to prove the theorem as
follows. The event $U_t=1$ with $t\ge 1$ is tantamount to the
condition that an integer $n\ge 1$ exists so that $T_n=t$, which in
particular yields $N_t=n$. Then, observing that $\prod_{i=1}^n
e^{v(s_i)}p(s_i)=e^{\ell t}\prod_{i=1}^n p_o(s_i)$ whenever
$s_1+\cdots+s_n=t$, for any Borel set $\mathcal{B}$ in $\Rl^d$ we have
\begin{eqnarray}
\nonumber
Z_t^c\cdot\,\prob_t^c\bigg[\frac{W_t}{t}\in \mathcal{B}\bigg]&=&\Ex\bigg[\mathds{1}_{\big\{\frac{W_t}{t}\in \mathcal{B}\big\}} U_te^{H_t}\bigg]\\
\nonumber
&=&\sum_{n\ge 1}\Ex\bigg[\mathds{1}_{\big\{\frac{1}{t}\sum_{i=1}^n f(S_i)\in \mathcal{B}\big\}} \mathds{1}_{\{T_n=t\}}e^{\sum_{i=1}^n v(S_i)}\bigg]\\
\nonumber
&=&\sum_{n\ge 1}\sum_{s_1\ge 1}\cdots\sum_{s_n\ge 1}\mathds{1}_{\big\{\frac{1}{t}\sum_{i=1}^n f(s_i)\in \mathcal{B}\big\}} \mathds{1}_{\{s_1+\cdots+s_n=t\}}
\prod_{i=1}^n e^{v(s_i)} p(s_i)\\
\nonumber
&=&e^{\ell t}\sum_{n\ge 1}\sum_{s_1\ge 1}\cdots\sum_{s_n\ge 1}\mathds{1}_{\big\{\frac{1}{t}\sum_{i=1}^n f(s_i)\in \mathcal{B}\big\}} \mathds{1}_{\{s_1+\cdots+s_n=t\}}
\prod_{i=1}^n p_o(s_i)\\
\nonumber
&=&e^{\ell t}\sum_{n\ge 1}\Ex_o\bigg[\mathds{1}_{\big\{\frac{1}{t}\sum_{i=1}^n f(S_i)\in \mathcal{B}\big\}} \mathds{1}_{\{T_n=t\}}\bigg]\\
\label{mod1}
&=&e^{\ell t}\,\Ex_o\Bigg[\mathds{1}_{\Big\{\frac{\sum_{i=1}^{N_t} f(S_i)}{\sum_{i=1}^{N_t}S_i}\in \mathcal{B}\Big\}} U_t\Bigg]\\
&\le&e^{\ell t}\,\prob_o\Bigg[\frac{\sum_{i=1}^{N_t}f(S_i)}{\sum_{i=1}^{N_t}S_i}\in \mathcal{B} \Bigg].
\label{mod2}
\end{eqnarray}
The identity (\ref{mod1}) with $\mathcal{B}=\Rl^d$ gives
$Z_t^c=e^{\ell t}\,\Ex_o[U_t]$, which shows that $e^{\ell t}\le
2\Ex_o[S_1]Z_t^c$ for all sufficiently large $t$ because of the limit
$\lim_{t\uparrow\infty}\Ex_o[U_t]=1/\Ex_o[S_1]$.  This way, using
$e^{\ell t}\le 2\Ex_o[S_1]Z_t^c$ in the bound (\ref{mod2}) specialized
to the closed set $\mathcal{B}:=\{w\in\Rl^d:\|w-\nu(0)\|\ge\delta\}$ and
dividing by $Z_t^c$, we find that for each $\delta>0$ and all
sufficiently large $t$
\begin{eqnarray}
\nonumber
  \prob_t^c\Bigg[\bigg\|\frac{W_t}{t}-\nu(0)\bigg\|\ge\delta\Bigg]\le
  2\Ex_o[S_1]\cdot\prob_o\Bigg[\bigg\|\frac{\sum_{i=1}^{N_t}f(S_i)}{\sum_{i=1}^{N_t}S_i}-\nu(0)\bigg\|\ge\delta \Bigg].
\end{eqnarray}
We obtain $\lim_{t\uparrow\infty}\,\prob_t^c[\|W_t/t-\nu(0)\|\ge
  \delta]=0$ from here thanks to (\ref{libconv}).

\ack
The author is grateful to Aernout van Enter for suggesting to include
the model by Fisher and Felderhof among renewal models of Statistical
Mechanics.

\end{document}